\documentclass[aps, prx, superscriptaddress, twocolumn, longbibliography, notitlepage]{revtex4-1}

\usepackage{amsfonts}
\usepackage[usenames, dvipsnames]{xcolor}
\usepackage{amsmath, amsthm, amssymb, dsfont}
\usepackage{enumerate}
\usepackage[english]{babel}
\usepackage{graphicx}
\usepackage[caption=false]{subfig}
\usepackage{bbm}
\usepackage[normalem]{ulem}
\usepackage{color}
\usepackage{hyperref}
\hypersetup{
	pdfnewwindow=true,
	colorlinks=true,
	linkcolor=red,
	citecolor=blue,
	filecolor=blue,
	urlcolor=blue
}

\colorlet{RED}{red}
\colorlet{BLUE}{blue}

\newcommand{\bra}[1]{\langle #1|}
\newcommand{\ket}[1]{|#1\rangle}

\newcommand{\ketbra}[2]{| #1 \rangle \langle #2 |}

\newcommand{\id}{\mathbbm{1}}
\newcommand{\bea}{\begin{eqnarray}}
\newcommand{\eea}{\end{eqnarray}}
\newcommand{\beaa}{\begin{eqnarray}\begin{aligned}}
\newcommand{\eeaa}{\end{aligned}\end{eqnarray}}
\newcommand{\cl}{\mathrm{cl}}

\newcommand{\ta}{\widetilde{a}}
\newcommand{\tm}{\widetilde{m}}
\newcommand{\tz}{\widetilde{z}}

\newcommand{\green}[1]{\textcolor[rgb]{0.,0.7,0.}{#1}}
\newcommand{\purple}[1]{\textcolor[rgb]{0.7,0.,0.5}{#1}}
\newcommand{\orange}[1]{\textcolor[rgb]{0.9,0.6,0}{#1}}

\DeclareMathOperator{\Tr}{Tr}

\begin{document}

\title{Minimizing back-action through entangled measurements}

\author{Kang-Da Wu}
\affiliation{CAS Key Laboratory of Quantum Information, University of Science and Technology of China, Hefei, 230026, People's Republic of China}
\affiliation{CAS Center For Excellence in Quantum Information and Quantum Physics, University of Science and Technology of China, Hefei, 230026, People's Republic of China}

\author{Elisa B\"{a}umer}
\affiliation{Institute for Theoretical Physics, ETH Zurich, 8093 Z\"{u}rich, Switzerland}

\author{Jun-Feng Tang}
\affiliation{CAS Key Laboratory of Quantum Information, University of Science and Technology of China, Hefei, 230026, People's Republic of China}
\affiliation{CAS Center For Excellence in Quantum Information and Quantum Physics, University of Science and Technology of China, Hefei, 230026, People's Republic of China}

\author{Karen V. Hovhannisyan}
\email{khovhann@ictp.it}
\affiliation{The Abdus Salam International Centre for Theoretical Physics (ICTP), Strada Costiera 11, 34151 Trieste, Italy}

\author{Mart\'{\i} Perarnau-Llobet}
\email{marti.perarnau@mpq.mpg.de}
\affiliation{Max-Planck-Institut f\"{u}r Quantenoptik, Hans-Kopfermann-Str. 1, D-85748 Garching, Germany}

\author{Guo-Yong Xiang}
\email{gyxiang@ustc.edu.cn}
\affiliation{CAS Key Laboratory of Quantum Information, University of Science and Technology of China, Hefei, 230026, People's Republic of China}
\affiliation{CAS Center For Excellence in Quantum Information and Quantum Physics, University of Science and Technology of China, Hefei, 230026, People's Republic of China}

\author{Chuan-Feng Li}
\affiliation{CAS Key Laboratory of Quantum Information, University of Science and Technology of China, Hefei, 230026, People's Republic of China}
\affiliation{CAS Center For Excellence in Quantum Information and Quantum Physics, University of Science and Technology of China, Hefei, 230026, People's Republic of China}

\author{Guang-Can Guo}
\affiliation{CAS Key Laboratory of Quantum Information, University of Science and Technology of China, Hefei, 230026, People's Republic of China}
\affiliation{CAS Center For Excellence in Quantum Information and Quantum Physics, University of Science and Technology of China, Hefei, 230026, People's Republic of China}

\begin{abstract}
	
	When an observable is measured on an evolving coherent quantum system twice, the first measurement generally alters the statistics of the second one, which is known as measurement back-action. We introduce, and push to its theoretical and experimental limits, a novel method of back-action evasion, whereby entangled collective measurements are performed on several copies of the system. This method is inspired by a similar idea designed for the problem of measuring quantum work [Perarnau-Llobet \textit{et al}., \href{https://doi.org/10.1103/PhysRevLett.118.070601}{Phys. Rev. Lett. \textbf{118}, 070601 (2017)}]. By utilizing entanglement as a resource, we show that the back-action can be extremely suppressed compared to all previous schemes. Importantly, the back-action can be eliminated in highly coherent processes.
	
\end{abstract}

\maketitle

\textit{Introduction}---The enthusiasm for probing and describing microscopic systems in the quantum domain motivates the search for ways to measure quantum processes, akin to standard measurements of operators. This task is notoriously challenging due to the fragility of quantum superpositions and invasiveness of measurements, epitomized in the quantum back-action (QBA) effect \cite{braginskii1975, caves1980, braginsky1996, wiseman2010}. Given two measurements separated by a certain time interval, there is QBA when the statistics of the second measurement are altered by the presence of the first measurement. It is mostly considered an unpleasant hindrance in experiments, and significant effort has been invested to design back-action-evading (a.k.a. quantum non-demolition) measurements \cite{braginskii1975, caves1980, wiseman1995, braginsky1996, kippenberg2008, wiseman2010, erhart2012, hacohen2016, moller2017, kono2018, pezze2018, wu2019, boulebnane2019}. Besides its fundamental interest, QBA also has important consequences for the definition of heat and work fluctuations in quantum thermodynamics \cite{allahverdyan2014, solinas2015, talkner2016, baumer2018, levy2019} and quantum transport \cite{hovhannisyan2019}. 

In its essence, QBA is an instance of the uncertainty principle \cite{peres1993}. Say, after measuring an observable $A$, the system's state, $\rho$, undergoes a unitary evolution $U$, generated by its internal Hamiltonian. Then, an observable $A'$ is measured. Viewing in the Heisenberg picture, we thus have two operators, $A$ and $A'_H = U^\dagger A' U$, simultaneously measured on the state $\rho$, and when $[A, A'_H] \neq 0$, these measurements are incompatible, both in the sense of the uncertainty relations \textit{\`{a} la} Heisenberg \cite{heisenberg1927, bialynicki1975, peres1993, maccone2014} and error-disturbance relations \textit{\`{a} la} Arthurs-Kelly \cite{arthurs1965, ishikawa1991, peres1993, ozawa2003, allahverdyan2013, branciard2013, erhart2012, busch2014, buscemi2014, perarnau2017}. Measurement incompatibility can also be described by constructing joint (quasi)probability distributions, with the condition that the marginal distributions be given by the Born rule \cite{wigner1932, terletsky1937, margenau1961, muckenheim1986, nazarov2003, hofer2017}. Quasiprobability distributions, generically arising in (sequential) weak measurement scenarios \cite{mitchison2007, avella2017, kim2018, shojaee2018, hovhannisyan2019, pfender2019, cujia2019, monroe2020}, necessarily feature negative probabilities \cite{margenau1961, cahill1969, busch1985, uffink1994, hartle2004, allahverdyan2014, potts2019} (which is related to contextuality \cite{spekkens2008, pusey2014, lostaglio2018}); joint distributions that do \textit{not} feature negative probabilities require going beyond the axiomatics of standard ``precise'' probability theory \cite{allahverdyan2018}. 

In this work, we will take a novel approach towards reducing QBA, by using collective measurements on multiple copies of the system of interest. The logic of our approach is as follows. When $[\rho, A] = 0$, the first measurement does not disturb the state, therefore, the second measurement is not affected by it. That is, there is \textit{no} QBA when the state is, in a sense, classical, and the joint probability is $p^{\cl}_{ij} = \Tr(\rho P_i P'_j) \geq 0$, where $P_i$ and $P'_j$ are the eigenprojectors of $A$ and $A_H'$, respectively ($A = \sum_i a_i P_i$, $A_H' = \sum_i a'_i P'_i$). This statistics coincides with that of the measure-evolve-measure protocol, widely used and discussed in quantum transport \cite{shelankov2003, nazarov2003, esposito2009, hovhannisyan2019} and thermodynamics \cite{jarzynski2004, talkner2007, esposito2009, perarnau2017nogo, levy2019}. 

When $[\rho, A] \neq 0$, the standard measure-evolve-measure protocol has QBA, and we wish to find new measurement schemes to reduce it (while agreeing with it for $[\rho, A] = 0$). However, it was recently shown that there cannot exist a state-independent generalized quantum measurement---a positive operator-valued measure (POVM)---that describes the joint statistics of $A$ and $A'_H$ and coincides with $p^{\cl}_{ij}$ when $\rho$ commutes with $A$ \cite{perarnau2017nogo}. Hence,
our idea is to consider schemes that are allowed to depend on the state, but do so as weakly as possible. To that end, we use measurements that can be realized as state-independent POVMs on several copies of the system \footnote{Only when viewed from the perspective of a single system, these depend on the state}. This  is motivated by a QBA-reduction technique proposed and implemented in Refs. \cite{perarnau2017nogo, wu2019}, for the problem of measuring quantum-mechanical work. There, it was also proven that the QBA cannot be removed no matter how many copies one takes; however, significant reduction can be observed even with two copies. Interestingly, the QBA reduction in Refs.~\cite{perarnau2017nogo, wu2019} was achieved by using   factorized  measurements (i.e., the POVM
elements  are in the form of a Kronecker product).

Here, focusing on the two-copy scenario, we go beyond these results and take this scheme to its extreme. By using entangled POVMs, we show that one can reduce the QBA even further, eliminating it completely in an extended parameter range. These enhancements are only possible due to the entangled nature of the measurements, as we also prove in Ref.~\footnote{See Supplemental Material at URL, which contains the mathematical details of the theoretical results and technical details of the experimental setup, and further references \cite{virmani2003, sanpera1998, verstraete2001}} that the proposal of Refs.~\cite{perarnau2017nogo, wu2019} is the optimal not only among factorized POVMs, but among all separable, i.e., possibly correlated but non-entangled, POVMs. We measure the QBA by comparing the statistics of $A'_H$ when $A$ is performed with that when $A$ is not performed. 
We determine the ultimate QBA-evading measurement on two copies, which turns out to be entangled. Somewhat counterintuitively, zero QBA can be achieved in a class of continuously parameterized highly coherent processes. We experimentally implement a class of such protocols in the discrete photonics regime, for observing minimal QBA in different quantum processes. Our results show the capability of achieving minimal QBA in our photonic setup with considerably low experimental imperfections, and demonstrate the significant advantage gained over non-entangled measurements.

\textit{Theoretical framework and results}---Let us formalize the discussion above. The statistics of $A$'s measurement is given by probabilities $p^{(1)}_i = \Tr(\rho P_i)$ and, if $A$ is not performed, the statistics of $A_H'$ is given by $p^{(2)}_j = \Tr(\rho P'_j)$. The desired goal is to find a $p^{(12)}_{ij} = \Tr(M^{ij} \rho)$, where $M^{(ij)}$ constitute a POVM \footnote{Namely, $M^{(ij)}$ are non-negative Hermitian operators such that $\sum_{ij} M^{(ij)} = \id$ \cite{wiseman2010}}, such that $\sum_j p^{(12)}_{ij} = p^{(1)}_i$ and $\sum_i p^{(12)}_{ij} = p^{(2)}_j$, for any state $\rho$. Had such state-independent POVM exist, we would say that $A$ and $A'_H$ are jointly measurable \cite{busch1985, uffink1994}. However, it is well-known that no such state-independent POVM exists that satisfies this goal for general $\rho$'s when $[A, A'_H] \neq 0$ \cite{busch1985, uffink1994, ozawa2003, branciard2013}. We deal with this impossibility in the following manner. First, we notice that, when $[\rho, H] = 0$,
\bea
p^{\cl}_{ij} = \Tr(\rho P_i P'_j) \equiv \Tr(\rho M_{\mathrm{TPM}}^{(ij)}),
\eea
where $M_{\mathrm{TPM}}^{(ij)} = P_i P'_j P_i$, describes a valid joint statistics and represents two projective measurements (hence the label ``TPM'') of $A$ and $A'_H$, performed one after another. Then, we recall that state-independent POVMs, that produce $p^{\cl}_{ij}$ whenever $[\rho, A] = 0$, essentially coincide with $M_{\mathrm{TPM}}^{(ij)}$ \cite{perarnau2017nogo}, meaning that there is no room for tweaking the QBA.

Therefore, inspired by an approach developed in the context of quantum thermodynamics \cite{perarnau2017nogo, wu2019}, we extend the class of POVMs by considering state-independent measurements on \textit{two} copies of the system: we take a general POVM, $\mathcal{M} = \{ M^{(ij)} \}$ ($M^{(ij)} \geq 0$ and $\sum_{ij} M^{(ij)} = \id$),
\bea \label{proideal}
p_{ij}[\mathcal{M}] = \Tr\big(\rho^{\otimes 2} M^{(ij)}\big),
\eea
and require it to satisfy the condition
\bea \hypertarget{Cone}
C_1 : \;\, \forall \rho \;\, \text{s.t.} \;\, [\rho, A] = 0, \;\, p_{ij}[\mathcal{M}] = p^{\cl}_{ij},
\eea
in other words, to reproduce $p^{\mathrm{cl}}_{ij}$ for any incoherent $\rho$. Furthermore, since this measurement is intended to describe the QBA effect, we also require the statistics produced by $\mathcal{M}$ for the first event (measurement of $A$) to coincide with $p^{(1)}_i$, for any $\rho$; i.e., $\mathcal{M}$ also satisfies
\bea \hypertarget{Ctwo}
C_2 : \;\, \forall \rho, \;\, \sum\nolimits_j p_{ij}[\mathcal{M}] = p^{(1)}_i .
\eea
Lastly, noticing that the set $C_1 \cap C_2$ has more than one element (as opposed to single-copy state-independent measurements, for which the above two requirements necessitate the measurement to coincide with $\mathcal{M}_{\mathrm{TPM}}$), we fulfill our main goal by finding $\mathcal{M}^{\mathrm{opt}} = \arg \min\limits_{\mathcal{M} \in C_1 \cap C_2} \mathcal{B}[\mathcal{M}]$, with $\mathcal{B}$ being the magnitude of the QBA itself:
\bea \label{bacac}
\mathcal{B}[\mathcal{M}] = \sqrt{\sum_j \left( p_j[\mathcal{M}] - p_j^{(2)} \right)^2},
\eea
where $p_j[\mathcal{M}] := \sum_i p_{ij}[\mathcal{M}]$ is the actual statistics of the second measurement. Put differently, we are looking for the POVM that delivers the correct statistics for classical states, always yields the Born rule for the first measurement, and minimizes the QBA for the second measurement (see also the discussion in Ref.~\cite{Note2}). Here, we quantified the QBA by the Euclidean distance between the probability distribution of the second measurement as produced by the joint POVM, $p_j$, and the probability distribution of the second measurement had the first measurement not taken place, $p^{(2)}_j$. In principle, one could choose other distance measures in \eqref{bacac}, but for our problem of interest---a two-level system---all of them will depend only on the ground-state probability and hence yield the same $\mathcal{M}^{\mathrm{opt}}$.

For simplicity, we will concentrate on qubits and consider the case of $A = A'$. Moreover, in the single-qubit Hilbert spaces, we will work in the eigenbasis of $A$, $\{\ket{0}, \ket{1}\}$, and, since constant shifts, and multiplication by constants, of $A$ do not affect the phenomena at hand, w.l.o.g, we can choose $A = \ketbra{1}{1}$. Below, we will sketch how $\mathcal{M}^{\mathrm{opt}}$ is calculated in this case, and refer the reader to Ref.~\cite{Note2} for full details. 

We start by considering requirement \hyperlink{Cone}{$C_1$}\newcommand{\Cone}{\hyperlink{Cone}{$C_1$}}. With our choice of operators, the TPM POVM elements will be $M_{\mathrm{TPM}}^{(ij)} = \ket{i}\bra{i} \cdot \vert \bra{j} U \ket{i}\vert^2$, resulting in
\begin{align} \label{eq:pcl}
p^{\cl}_{ij} = \rho_{ii} |U_{ji}|^2,
\end{align}
where $i, j$ are $0$ or $1$ and $\rho_{ii} := \bra{i} \rho \ket{i}$ describes the probability to start in the state $\ket{i}$ and $|U_{ji}|^2 := \left|\bra{j} U \ket{i} \right|^2$ the transition probability from $\ket{i}$ to $\ket{j}$. Now, taking into account that all four matrices $M^{(ij)}$ do not depend on the state, Eq.~\eqref{eq:pcl} fixes all diagonal elements of these matrices up to some constants $a$ and $\widetilde{a}$ (see \cite{Note2} for the explicit construction). In particular, one diagonal element will always be zero, so the matrices are at most of rank three.

In order to minimize the QBA, we have to divide the construction and analysis into two cases. First, \textit{highly coherent} unitary operators, that is, evolutions that produce a lot of coherence given a classical state: $|U_{10}|^2 \in \left[\frac{1}{3}, \frac{2}{3}\right]$. Second, \textit{slightly coherent} unitary operations: $|U_{10}|^2 \in \left(0, \frac{1}{3}\right) \cup \left(\frac{2}{3}, 1 \right)$. Interestingly, for the first case one can construct a POVM with zero QBA, while it \textit{cannot} be reduced to zero for the second case.

In the case of highly coherent unitaries, we can write the optimal POVM as four rank-two matrices that are each a convex combination of $\ket{\psi^-}=(\ket{01}-\ket{10})/\sqrt{2}$ and a non-trivial pure entangled state. Note that this POVM also satisfies requirement \hyperlink{Ctwo}{$C_2$}\newcommand{\Ctwo}{\hyperlink{Ctwo}{$C_2$}}, i.e., $\Tr[(M^{(00)} + M^{(01)}) \rho^{\otimes 2}] = \rho_{00}$. In the case of slightly coherent unitaries, the optimal POVM contains two rank-three matrices and two rank-one matrices. In this case, there is always some back-action proportional to the initial coherence $\rho_{01}$ (see Ref.~\cite{Note2}). 

In both cases, the optimal POVM are entangled, and one may wonder to which extent such quantum correlations are needed. We also show that the separable collective measurement previously proposed in Refs.~\cite{perarnau2017nogo, wu2019} is in fact the optimal non-entangled POVM. Hence, any advantage reported in this work is possible due to the entangled nature of the measurement (see Ref.~\cite{Note2} for details).

\begin{figure}[htp]
	\label{fig:exp}
	\includegraphics[scale=0.11]{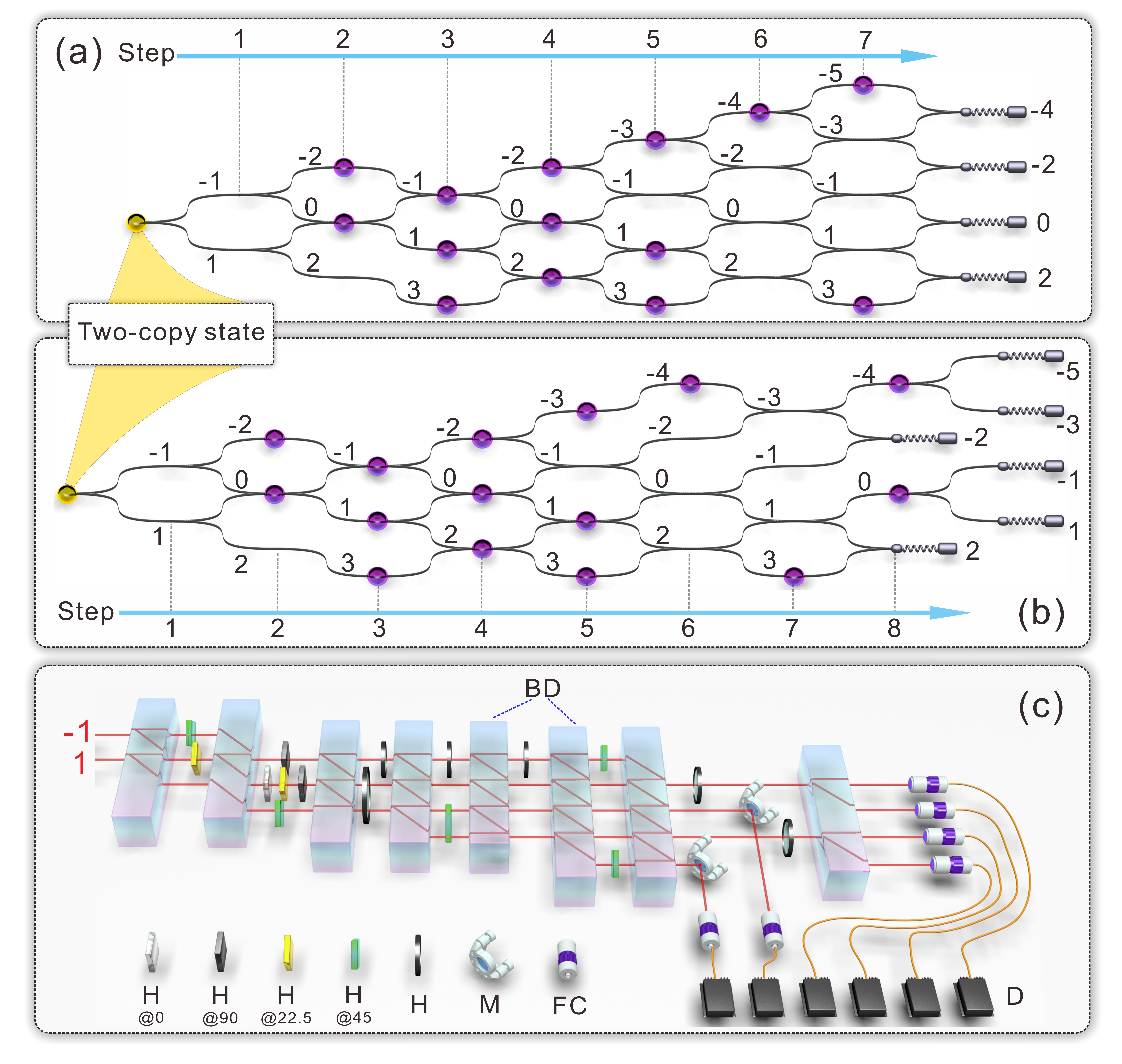}
	\caption{\label{fig:exp} \textbf{Experimental setup for optimal entangled collective measurement.} \textbf{(a)} and \textbf{(b)} detail the quantum walk setup for both, the perfect and imperfect case, respectively. The orange node represents the two-copy state preparation, which is independent of the measurement, whose implementation is shown in detail in the Supplementary Materials. A purple node represents a coin operator depending on a coordinate $(p, t)$ in the position-time space. In \textbf{(c)}, the whole experimental setup is shown, where the two cases can be simulated with one experimental setup. The details will be given in the Supplemental Material. Key to components: BD---beam displacer; H---half-wave plate; M---mirror; D---single photon detector.
	}
\end{figure}

\begin{figure}[htp]
	\label{fig:result1}
	\includegraphics[scale=0.165]{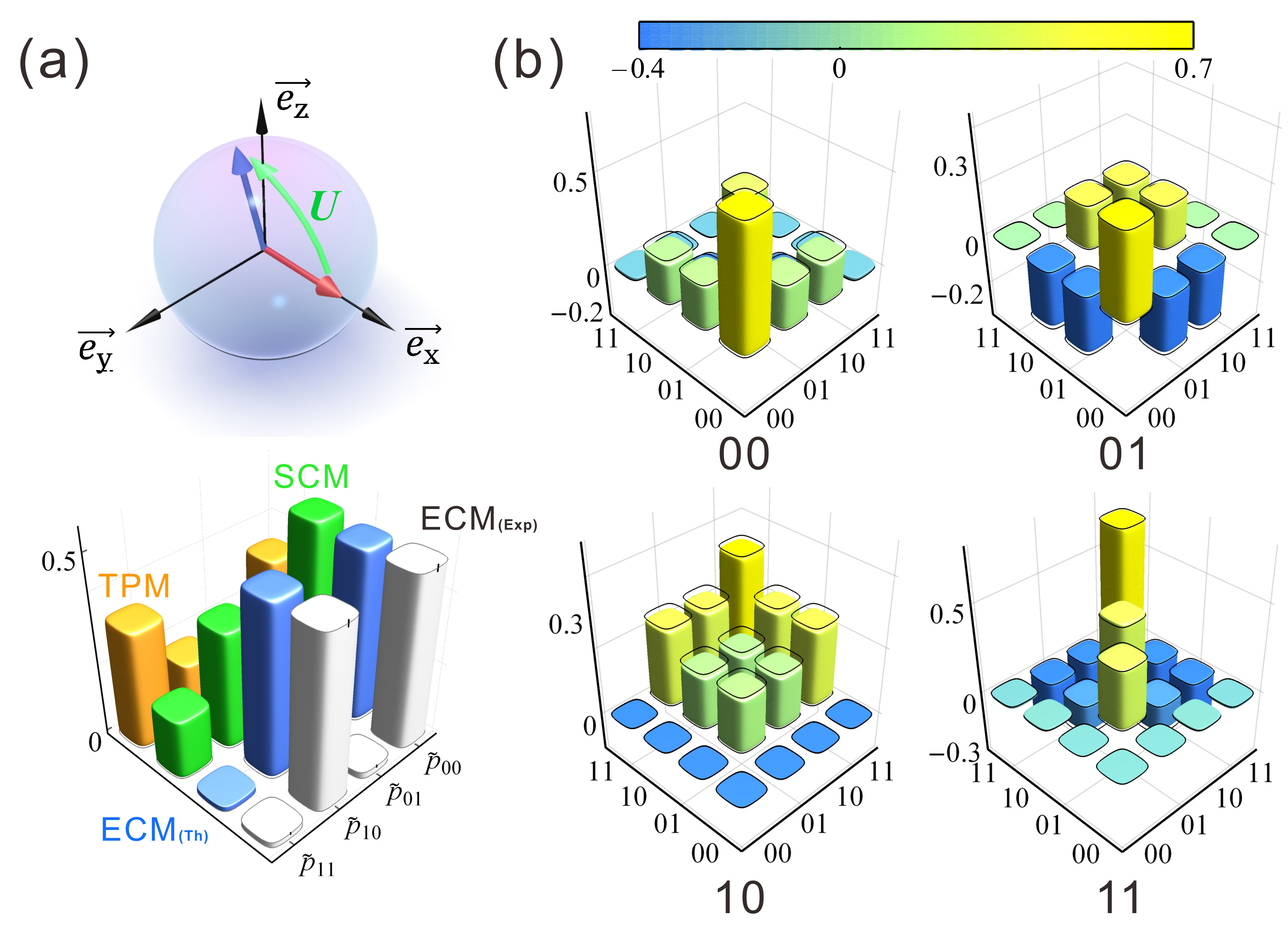}
	\caption{\label{fig:result1} \textbf{Experimental results for $U_{01} = \sqrt{1/3}$. }\textbf{(a)} Bloch representation for the ideal dynamics from input state to final state is shown with experimentally obtained $\widetilde{p}_{ij}$ (in white cubes), simulated values for the entangled collective measurements (blue; labeled as ``ECM''), optimal separable collective measurements (green; labeled as ``SCM'') and TPM (orange). \textbf{(b)} Real part of the experimentally reconstructed entangled POVMs, shown as colored long cubes with ideal values in transparent black edged cubes.  
		The average fidelity reads $0.985$. 
	}
\end{figure}

\begin{figure}[htp]
	\includegraphics[scale=0.24]{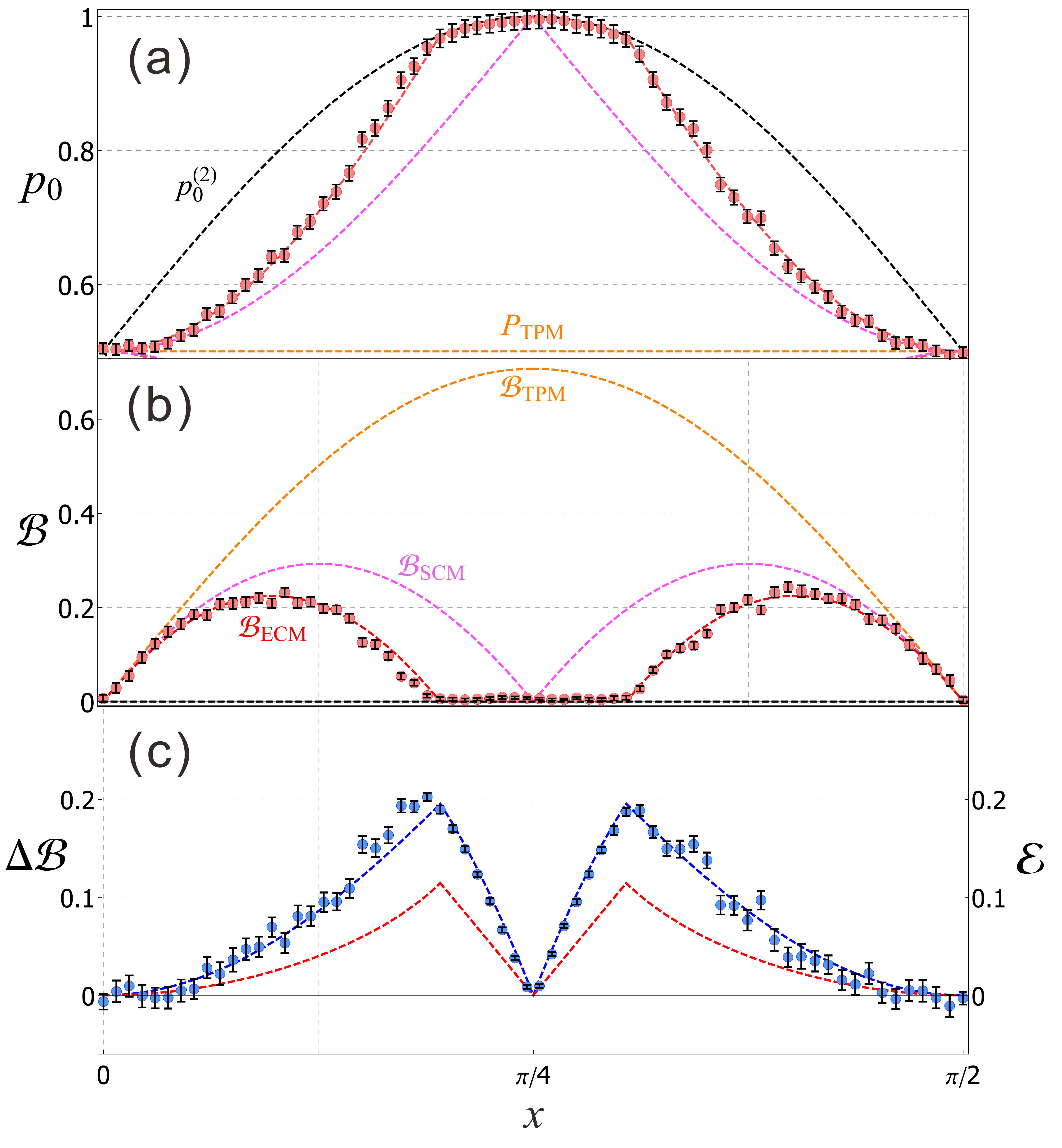}
	\caption{\label{fig:result2} \textbf{Experimental results for various processes.} \textbf{(a)} Experimental probabilities for final state being $\ket{0}$ ($\widetilde{p}_0 \geq 1/2$), plotted against $x$, are shown as red disks. Also, we show the simulated values for TPM (orange), separable collective measurements (pink),
		new entangled collective measurement (red)
		and the ideal case (black). \textbf{(b)} Experimentally obtained back-action, defined, in analogy with Eq.~\eqref{bacac}, as $\widetilde{\mathcal{B}} = \sqrt{\sum_j \left(\widetilde{p}_j - p^{(2)}_j \right)^2}$, plotted against $x$ as red disks. Simulated values for TPM (orange), separable collective measurements (pink;  labeled as ``$\mathcal{B}_{\mathrm{SCM}}$''), new entangled collective measurement (red; labeled as ``$\mathcal{B}_{\mathrm{ECM}}$'') and the ideal case (black) are also shown as dashed lines. \textbf{(c)} Experimental and theoretical gains (respectively, blue disks and line) in QBA reduction of entangled measurements over separable ones, $\Delta \mathcal{B} = \mathcal{B}_{\mathrm{SCM}} - \mathcal{B}_{\mathrm{ECM}}$, and the entanglement, $\mathcal{E}$, (red line) of the optimal POVM. 
	}
\end{figure}

\begin{figure}[htp]
	\label{fig:result3}
	\includegraphics[scale=0.24]{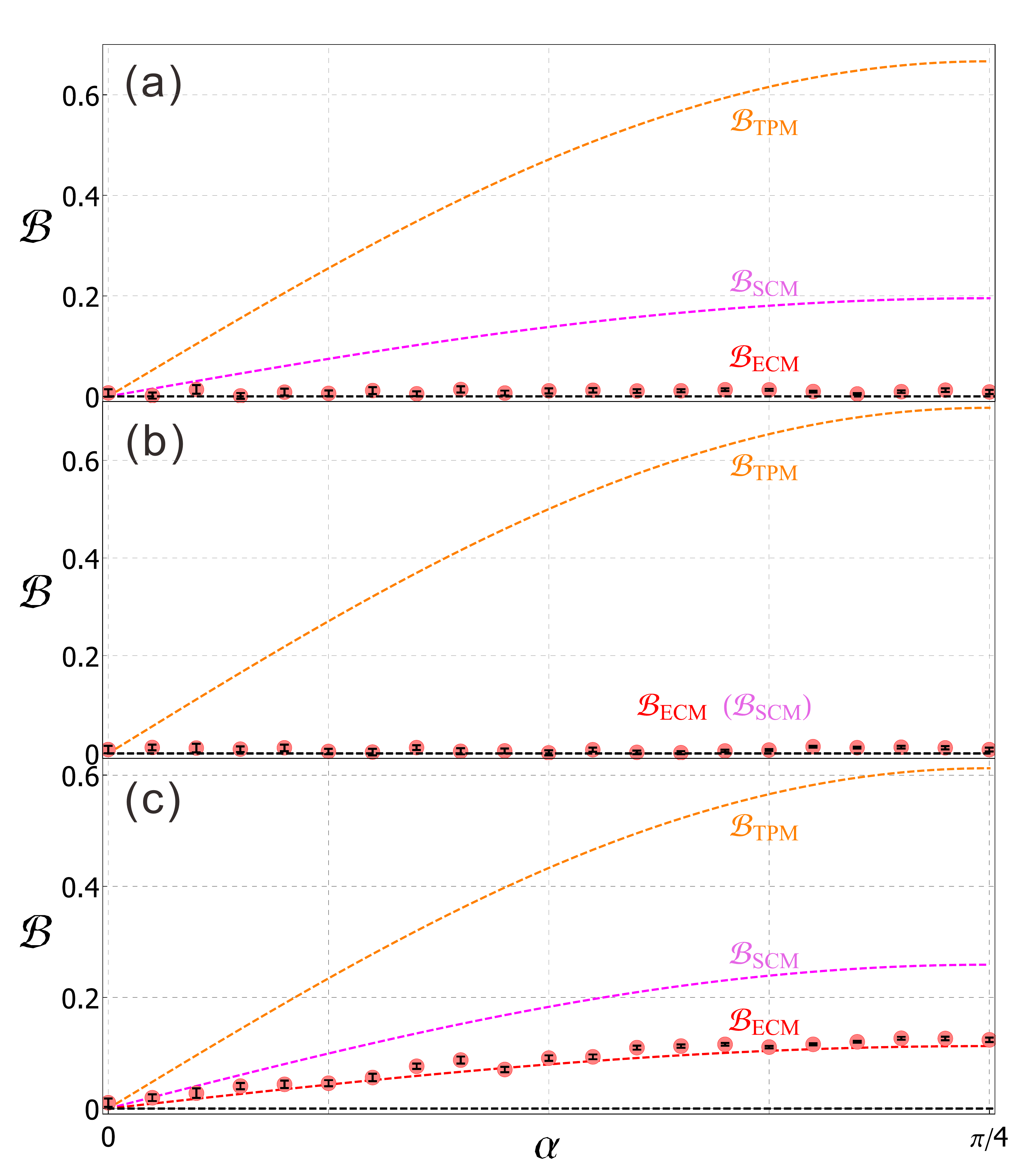}
	\caption{\label{fig:result3} \textbf{Experimental results for various inputs and fixed processes.} We choose three fixed processes for our experiments, $U_{01} = 1/\sqrt{3}, \, 1/\sqrt{2}, \, 1/2$ for \textbf{(a)}, \textbf{(b)}, \textbf{(c)}, respectively. Experimentally obtained back-actions are plotted against $\alpha$, with simulated values for TPM (orange), separable collective measurements (pink;  labeled as ``$\mathcal{B}_{\mathrm{SCM}}$''), new entangled collective measurement (red;  labeled as ``$\mathcal{B}_{\mathrm{ECM}}$'') and the ideal case (black). 
	}
\end{figure}

\textit{Experiment}---We now experimentally implement the minimally invasive entangled POVMs derived above.

Fig.~\ref{fig:exp} details both the conceptual design of our quantum walk and the corresponding experimental setup. For state preparation, we make use of multiple degrees of freedom of a down-converted single photon \cite{takeuchi2001} (spatial mode and polarization). Specifically, we encode the first copy in the spatial degree of freedom, and the second copy in the polarization degree of freedom \cite{wu2019, hou2018}, preparing the state $\ket{\Phi}\otimes\ket{\Phi}$, where $\ket{\Phi} = \cos \alpha\ket{0} + \sin\alpha\ket{1}$ (see Ref.~\cite{Note2} for details).

Experimentally demonstrating the highly non-trivial entangled collective measurements is much more challenging than the separable measurements previously reported in Ref.~\cite{wu2019}. The realization of the POVMs can be embedded in the dynamics of a one-dimensional discrete quantum walk \cite{xiao2017, hou2018, xue2015, zhao2015} containing two degrees of freedom: the walker position, $q$, and a coin state, $c$. The coin qubit and the walker in positions, respectively, $1$ and $-1$ are taken as the two-qubit system of interest, while the other positions of the walker are regarded as an auxiliary system. By engineering the coin operators followed by measuring the walker position after certain steps, a POVM can be deterministically realized with desired outcome at each position. In Figs.~\ref{fig:exp}(a) and (b), the diagram of quantum walks for two cases are shown. The perfect QBA elimination in the case of highly coherent processes can be realized via a $7$ step quantum walk, while the slightly coherent case needs $8$ steps.

Our all-optical setup is capable of realizing the POVMs for both cases described above. 
Fig.~\ref{fig:exp}(c) shows the experimental setup for realizing these POVMs. The whole setup is based on a single-photon interferometer. Robust interference between photons in different modes is required to ensure the high-fidelity-implementation of the measurement operators. Details are given in Ref.~\cite{Note2}.

A big entanglement-induced advantage over previous schemes (TPM and the optimal separable collective measurement \cite{perarnau2017nogo}) emerges when $|U_{01}|^2 = 1/3$. In particular, Fig.~\ref{fig:result1}(a) shows an initial state (red) and final state (blue) in Bloch representation: a pure state $\ket{+}$ is initialized and evolves into $\sqrt{1/3} \ket{+} +\sqrt{2/3}\ket{-}$, where $\ket{\pm}=(\ket{0} \pm \ket{1})/\sqrt{2}$. The experimentally obtained $\widetilde{p}_{ij}$ (where the $\sim$ above a quantity means that is evaluated experimentally) from entangled measurement are presented in Fig.~\ref{fig:result1}(a) as white cubes, while the ideal values given by Eq.~(\ref{proideal}) are plotted as blue cubes. The expected results obtained from previous schemes are also shown as green (for separable collective measurement) and orange (for TPM) cubes. Fig.~\ref{fig:result1}(b) shows the experimentally reconstructed POVMs, with an average fidelity up to $0.985$. The fidelity between two POVMs can be defined as $\mathcal{F}(M, M') = \left(\sum_i w_i \sqrt{\mathcal{F}_i}\right)^2$, where $w_i = \sqrt{\Tr(M_i) \Tr(M'_i)}/d$, and $\mathcal{F}_i$ is the fidelity between $\mu_i := \frac{M_i}{\Tr(M_i)}$ and $\mu'_i := \frac{M'_i}{\Tr(M'_i)}$~\cite{hou2018}. 

Fig.~\ref{fig:result2} shows the experimental reduction of QBA for a family of unitaries parameterized as $U_{00} = - U_{11} = \cos x$ and $U_{01} = U_{10} = \sin x$, where $x$ determines the ``degree of coherence'' of $U$. Fig.~\ref{fig:result3} shows a similar analysis for different input states and a fixed process. In both figures, the QBA is quantified through Eq.~\eqref{bacac}.


To observe the divergence between these methods, we first fix the initial state to be the maximally coherent state $\ket{+}$ and implement $67$ entangled POVMs so that both highly coherent and slightly coherent processes are included. The results are shown in Fig.~\ref{fig:result2}: (a) shows the probabilities $p_0$ of ending states under various protocols, the experimental values for new entangled POVMs are shown as red disks; (b) shows the experimentally observed QBA for the new POVMs and the simulated values for QBA of previous schemes. The new entangled POVMs (red points) can, within tolerable experimental imperfection, produce probabilities for final states that are closest to the ideal case (black lines), resulting in a significant decrease in QBA compared to the separable measurement. Fig.~\ref{fig:result2}(c) illustrates the fact that any advantage reported in this work is possible due to the entangled nature of the measurement, by showing a comparison between the maximal gain in QBA reduction entangled measurements offer over separable ones and the value of entanglement, as measured by the negativity~\cite{horodecki2009}, of the measurement delivering that gain.

We also experimentally test the performance of the new scheme on different inputs $\ket{\Phi}$ (parameterized by $\alpha$) with fixed $x$. Specifically, we choose $U_{01} =\sqrt{1/3}, \sqrt{1/2}, 1/2$, and the coherence of inputs, quantified as $\sin (2 \alpha)$, varies from $0$ to $1$. The experimentally observed QBA for the entangled POVMs, and the simulated values for all schemes, are shown in Fig.~\ref{fig:result3}, where we can see that the experimentally observed QBA is almost zero for two highly coherent processes and is significantly advantageous over the separable ones for slightly coherent process. The error bars in all figures are estimated by Monte Carlo simulations~\cite{Note2}.

\textit{Conclusions}---In this work, we demonstrated that harnessing non-trivial entangled measurements can significantly minimize the QBA. We devised the optimal measurement for reducing the QBA given two copies of a qubit, which turns out to be an entangled measurement. We also proved that the optimal non-entangled measurement coincides with the previous proposal of \cite{perarnau2017nogo, wu2019}. More specifically, we show that for highly coherent unitary operations the QBA can, in principle, be eliminated. While such an elimination is impossible for slightly coherent unitaries, our new scheme offers a considerable advantage over those using only separable measurements \cite{perarnau2017nogo, wu2019}.

Our experimental implementation of the optimal measurement demonstrates substantial entanglement-based advantages over previous schemes ~\cite{perarnau2017nogo, wu2019}. Our design can be realized with high average fidelity reaching up to $0.985$, which results in a good agreement between theoretical predictions and experimental observations.

It is worth emphasizing that the measurement step of our experiment is completely independent of the preparation step. In fact, other preparation procedures can be used. For example, by using quantum state joining \cite{vitelli2013joining}, one can convert the joint state of two photons (which could be generated, e.g., in an atom-photon interface \cite{blinov2004observation, wilk2007single, stute2012tunable}) into a hybrid encoded single-photon state. Our experimental setup thus demonstrates the feasibility of in-depth investigation of fundamental principles in quantum mechanics within a variety of current technologies.

\section*{Acknowledgement}The work at the University of Science and Technology of China is supported by the National Key Research and Development Program of China (Nos.2017YFA0304100 and 2018YFA0306400), the National Natural Science Foundation of China (Grants Nos. 61905234, 11974335, 11574291, and 11774334), the Key Research Program of Frontier Sciences, CAS (Grant No. QYZDYSSW-SLH003) and the Fundamental Research Funds for the Central Universities (Grant No. WK2470000026). E. B. acknowledges the support from the Swiss National Science Foundation via the NCCR QSIT and project No. 200020 165843 and the EU COST Action MP1209 on Thermodynamics in the Quantum Regime.

%

\newpage
\begin{widetext}
\appendix
\section{Remarks on \textcolor{red}{$C_1$} and \textcolor{red}{$C_2$}}
\label{app:remarks}

In order to better understand the necessity of imposing both requirements \textcolor{red}{$C_1$} and \textcolor{red}{$C_2$} in the main text, let us make the following two observations.

First, we note that, for a given state, it would be possible to construct a POVM that does not have any back-action for all unitaries, at the price of dropping \textcolor{red}{$C_1$}; the POVM will have to depend on the state, otherwise, $A$ and $A'_H$ would be jointly measurable for any $U$, which is impossible \cite{busch1985, uffink1994} (Refs. [39, 40] in the main text). With such a POVM, one would be able to recover the correct initial and final statistics, however, given a classical state, one would not recover the classical fluctuations. To give an example, if one would start in the maximally mixed state, i.e., with probability $1/2$ in state $\ket{0}$ and with probability $1/2$ in state $\ket{1}$, and would apply a bit flip, then classically we would expect to have probability $1/2$ to go from $\ket{0}$ to $\ket{1}$ and  probability $1/2$ to go from $\ket{1}$ to $\ket{0}$. However, a POVM constructed in a way to give the correct initial and final distributions, would instead yield the fluctuations $p_{00} = p_{01} = p_{10} = p_{11} = 1/4$. This incompatibility (at least for one copy) is expected due to the no-go theorem in Ref.~\cite{perarnau2017nogo} (Ref. [51] in the main text).

Second, if we were to drop the positivity constraint, \textcolor{red}{$C_2$}, we could construct a set of matrices $M^{(ij)}$ that add up to identity, yield the correct statistics for classical states, and have no back-action \cite{terletsky1937, margenau1961, muckenheim1986} (Refs. [33, 34, 35] in the main text). In addition, $p_{ij} = \Tr[M^{(ij)} \rho^{\otimes 2}] \geq 0$ would even yield a valid probability distribution. However, some of the matrices $M^{(ij)}$ would not be positive-semidefinite, meaning that, while the probabilities $p_{ij}$ would be non-negative when applying $M^{(ij)}$ to states of the form $\rho \otimes \rho$, there would necessarily exist certain two-qubit states for which $M^{(ij)}$ would produce negative probabilities. This means that such a distribution can not be directly measured in an experimental set-up (it could however be obtained, e.g., via tomography). 

These points illustrate that, although in some specific instances one could produce a probability distribution $p_{ij}$ whose appropriate marginals would describe the statistics of $A$ and $A'_H$, such schemes would, at core, be unphysical.

\section{QBA-minimizing POVMs} \label{app:newconstruction}

Let us start the construction of the optimal entangled POVM by considering requirement \textcolor{red}{$C_1$}. We therefore require
\begin{align}
\forall \rho = D_H (\rho), \quad i, j \in \{0, 1\}: \qquad p_{ij}^{\mathrm{cl}} \overset{!}{=} p_{ij} = \Tr [\rho^{\otimes 2} M^{(ij)}] = \rho_{00}^2 M^{(ij)}_{00} + \rho_{00} \rho_{11} \left( M_{11}^{(ij)} +M_{22}^{(ij)} \right) + \rho_{11}^2 M_{33}^{(ij)}, 
\end{align}
where $M_{kl}^{(ij)}:= \bra{k}M^{(ij)} \ket{l}$.
As the POVM elements cannot depend on $\rho$, using 
\begin{align} \label{eq:pcl}
p^{\text{cl}}_{ij} = \rho_{ii} |U_{ji}|^2,
\end{align}
for the diagonal entries we find
\begin{align}
\green{|U_{00}|^2} &= M_{00}^{(00)} = M_{11}^{(00)} + M_{22}^{(00)}
\\
&=  M_{33}^{(11)} = M_{11}^{(11)} + M_{22}^{(11)},
\\
\green{|U_{10}|^2} &= M_{00}^{(01)} = M_{11}^{(01)} + M_{22}^{(01)}
\\
&=  M_{33}^{(10)} = M_{11}^{(10)} + M_{22}^{(10)},
\\
\green{0} &= M_{33}^{(00)} + M_{33}^{(01)} +M_{00}^{(01)} + M_{00}^{(10)}.
\end{align}
Next, using the fact that these form a POVM, i.e., $\sum_{i, j} M^{(ij)} = \id$ and $M^{(ij)}\geq 0$, $\forall i, j$, we end up with the following construction:
\beaa \label{construction}
M^{(00)} = 
\begin{pmatrix}
	\green{|U_{11}|^2 } & \purple{-m_1^*} & \purple{-m_2^*} & 0\\
	\purple{-m_1} & \green{\ta|U_{11}|^2} & \orange{-z'^*} & 0\\
	\purple{-m_2} & \orange{-z'} & \green{(1-\ta) |U_{11}|^2 } & 0\\
	0 & 0 & 0 & \green{0}\\
\end{pmatrix},
\qquad  \qquad
M^{(01)} = 
\begin{pmatrix}
	\green{|U_{10}|^2 } & \purple{m_1^*} & \purple{m_2^*} & 0\\
	\purple{m_1} & \green{a|U_{10}|^2} & \orange{z^*} & 0\\
	\purple{m_2} & \orange{z} & \green{(1-a) |U_{10}|^2 } & 0\\
	0 & 0 & 0 & \green{0}\\
\end{pmatrix},
\\
M^{(10)} = 
\begin{pmatrix}
	\green{0} & 0 & 0 & 0\\
	0 & \green{(1-a) |U_{10}|^2} & \orange{\tz^*} & \purple{- \tm_2^*} \\
	0 & \orange{\tz} & \green{a |U_{10}|^2} & \purple{- \tm_1^*} \\
	0 & \purple{- \tm_2} & \purple{- \tm_1} & \green{|U_{10}|^2} \\
\end{pmatrix}, 
\qquad \qquad
M^{(11)} = 
\begin{pmatrix}
	\green{0} & 0 & 0 & 0\\
	0 & \green{(1-\ta) |U_{11}|^2} & \orange{-\tz'^*} & \purple{\tm_2^*} \\
	0 & \orange{-\tz'} & \green{\ta |U_{11}|^2} & \purple{\tm_1^*} \\
	0 & \purple{\tm_2} & \purple{\tm_1} & \green{|U_{11}|^2} \\
\end{pmatrix},
\eeaa
with $\orange{z + \tz = z' + \tz'}$. Now we optimize according to 
\bea \label{optmeas}
\mathcal{M}^{\mathrm{opt}} = \arg \min_{\mathcal{M} \in C_1 \cap C_2} \max_{j} \left\vert \sum\nolimits_i p_{ij}[\mathcal{M}] - p_j^{(2)} \right\vert.
\eea
The probability distribution of the final unmeasured state in the final basis is given by
\begin{align}
p_0^{(2)} &:= \bra{0} U \rho U^\dag \ket{0} = \green{ |U_{11}|^2\rho_{00}+|U_{10}|^2\rho_{11} } - \purple{ 2 Re [\rho_{01} U_{10}^\ast U_{11} ]} \label{unmeas1}\\
p_1^{(2)} &:= \bra{1} U \rho U^\dag \ket{1} = \green{|U_{10}|^2\rho_{00}+|U_{11}|^2\rho_{11}} + \purple{2 Re[\rho_{01} U_{10}^\ast U_{11} ]}.\label{unmeas2}
\end{align}
This should be as close as possible to the probability distribution of the measured state, 
\begin{align}
p_{00} + p_{10} &= \Tr[\rho^{\otimes 2} (M^{(00)}+M^{(10)})] \nonumber \\
&= \green{|U_{11}|^2\rho_{00}+|U_{10}|^2\rho_{11}} - \purple{2 Re[\rho_{01}\left( \rho_{00} (m_1+m_2) + (1-\rho_{00}) (\tilde{m}_1+\tilde{m}_2) \right)]} + \orange{2Re[|\rho_{01}|^2(\tz - z')]} \label{meas1}  \\
p_{01} + p_{11} &= \Tr[\rho^{\otimes 2} (M^{(01)}+M^{(11)})] \nonumber\\
&=  \green{|U_{10}|^2\rho_{00}+|U_{11}|^2\rho_{11}} + \purple{2 Re[\rho_{01}\left( \rho_{00} (m_1+m_2) + (1-\rho_{00}) (\tilde{m}_1+\tilde{m}_2) \right)]} +\orange{2Re[|\rho_{01}|^2(z - \tz')]}.\label{meas2}
\end{align}
We can see that the first two (green) terms of Eqs.~\eqref{meas1} and \eqref{meas2} correspond to those of Eqs.~ \eqref{unmeas1} and \eqref{unmeas2}. As the last (orange) part in Eqs.~\eqref{meas1} and \eqref{meas2} is proportional to $|\rho_{01}|^2$, we require that $\orange{\tz = z'}$ and $\orange{\tz' = z}$.
Ideally, we would now like to set $\purple{m_1 + m_2 = \tilde{m}_1 + \tilde{m}_2 = U_{10} U_{11}^*}$, such that $p_0^{(2)} = p_{00} + p_{10}$ and $p_1^{(2)} = p_{01} + p_{11}$. However, due to positivity of our POVM matrices, we get some constraints, among others:
\begin{align}
0&\leq a\leq 1, \quad 0 \leq \ta \leq 1
\\ \label{m1cond}
|m_1|&\leq \min\{ |U_{10}|^2\sqrt{a}, |U_{11}|^2\sqrt{\ta}\} \qquad
\\ \label{m2cond}
|m_2|&\leq \min\{ |U_{10}|^2\sqrt{1-a}, |U_{11}|^2\sqrt{1-\ta}\}
\\ \label{m1tcond}
|\tilde{m}_1|&\leq \min\{ |U_{10}|^2\sqrt{a}, |U_{11}|^2\sqrt{\ta}\} \qquad
\\ \label{m2tcond}
|\tilde{m}_2|&\leq \min\{ |U_{10}|^2\sqrt{1-a}, |U_{11}|^2\sqrt{1-\ta}\}, 
\end{align}
which implies that $\purple{|m_1 + m_2| \leq \min\{|U_{10}|^2(\sqrt{a}+\sqrt{1-a}), |U_{11}|^2(\sqrt{\ta}+\sqrt{1-\ta})\}\leq \sqrt{2} \min\{|U_{10}|^2, |U_{11}|^2\} }$. Thus, for slightly coherent unitaries where $\sqrt{2} \min\{|U_{10}|^2, |U_{11}|^2\}\leq |U_{10} U_{11}|$, we should set $\green{a = \ta = \frac{1}{2}}$ in order to maximize $|m_1+m_2|$. We will see that for highly coherent unitaries this is also a good choice, as for those we can thereby construct a POVM that has no back-action at all.

\subsection{Highly coherent unitaries}

First we look at the construction of the POVMs in the perfect case of highly coherent unitaries, i.e. where 
\begin{align}
\sqrt{2} \min\{|U_{10}|^2, |U_{11}|^2\}\geq |U_{10} U_{11}| \quad \Leftrightarrow \quad \frac{1}{3} \leq |U_{10}|^2 \leq \frac{2}{3}.
\label{case1}
\end{align}
In this case we can set $m_1=m_2=\tilde{m}_1=\tilde{m}_2=\frac{U_{10}U_{11}^*}{2}$ and thereby reach zero back-action. Now we still need to find some $z = \tz'$ and $\tz = z'$, such that all the POVM matrices are positive-semidefinite. Taking $z$ to be real and calculating the determinants of the four matrices, we get the following conditions:
\begin{align}
\left(z - \frac{|U_{10}|^2}{2}\right)\left(z - \frac{|U_{11}|^2 - |U_{10}|^2}{2}\right) &\leq 0
\label{firstcond}\\
\left(\tz - \frac{|U_{10}|^2}{2}\right) \left(\tz-\frac{|U_{11}|^2-|U_{10}|^2}{2}\right) &\leq 0 \label{secondcond}\\
\left(z' + \frac{|U_{11}|^2}{2}\right) \left(z' + \frac{|U_{10}|^2-|U_{11}|^2}{2}\right) &\leq 0 \label{thirdcond}\\
\left(\tz' + \frac{|U_{11}|^2}{2}\right) \left(\tz' + \frac{|U_{10}|^2-|U_{11}|^2}{2}\right) &\leq 0
\end{align}
We can immediately see that we need to set $\orange{z = \tz = z' = \tz' = \frac{|U_{11}|^2 - |U_{10}|^2}{2}}$ in order to have all POVM matrices positive-semidefinite. Note that since now $z = z'$, it follows directly that also the initial probabilities are conserved, 
\begin{align} \label{initialprob}
p_{00} + p_{01} = \Tr[(M^{(00)} + M^{(01)})\rho^{\otimes 2}] = \rho_{00} + 2 Re [z - z'] |\rho_{01}|^2 = \rho_{00} \nonumber\\
p_{10} + p_{11} = \Tr[(M^{(10)} + M^{(11)})\rho^{\otimes 2}] = \rho_{11} - 2 Re [z - z'] |\rho_{01}|^2 = \rho_{11}.
\end{align}

Characterizing the unitary by $U = \sigma_{\mathbbm{z}} \cos(x) + \sigma_{\mathbbm{x}} \sin(x)$, we can substitute $\sin(x)=U_{10}$ and $\cos(x)=-U_{11}$ and the POVM matrices in this case (i.e., for $\frac{1}{3}\leq\sin(x)^2\leq \frac{2}{3}$) are therefore given by
\begin{align}
M^{(00)}&=\begin{pmatrix}
\cos(x)^2 & \frac{\cos(x)\sin(x)}{2} & \frac{\cos(x)\sin(x)}{2} & 0\\
\frac{\cos(x)\sin(x)}{2} & \frac{\cos(x)^2}{2} & \frac{\sin(x)^2-\cos(x)^2}{2} & 0\\
\frac{\cos(x)\sin(x)}{2} &  \frac{\sin(x)^2-\cos(x)^2}{2} & \frac{\cos(x)^2}{2} & 0\\
0 & 0 & 0 & 0
\end{pmatrix}
= \begin{pmatrix}
\frac{1+\cos(2x)}{2}&  \frac{\sin(2x)}{4}  & \frac{\sin(2x)}{4} & 0\\
\frac{\sin(2x)}{4}  &\frac{1+\cos(2x)}{4} & -\frac{cos(2x)}{2}& 0\\
\frac{\sin(2x)}{4} &  -\frac{cos(2x)}{2}  &\frac{1+\cos(2x)}{4} & 0\\
0 & 0 & 0 & 0
\end{pmatrix}\\
M^{(01)}&=\begin{pmatrix}
\sin(x)^2 & -\frac{\cos(x)\sin(x)}{2} & -\frac{\cos(x)\sin(x)}{2} & 0\\
-\frac{\cos(x)\sin(x)}{2} & \frac{\sin(x)^2}{2} & \frac{\cos(x)^2-\sin(x)^2}{2} & 0\\
-\frac{\cos(x)\sin(x)}{2} &  \frac{\cos(x)^2-\sin(x)^2}{2} & \frac{\sin(x)^2}{2} & 0\\
0 & 0 & 0 & 0
\end{pmatrix}
= \begin{pmatrix}
\frac{1-\cos(2x)}{2}& -\frac{\sin(2x)}{4}  & -\frac{\sin(2x)}{4} & 0\\
-\frac{\sin(2x)}{4}  &\frac{1-\cos(2x)}{4} & \frac{cos(2x)}{2}& 0\\
-\frac{\sin(2x)}{4} &  \frac{cos(2x)}{2}  &\frac{1-\cos(2x)}{4} & 0\\
0 & 0 & 0 & 0
\end{pmatrix}\\
M^{(10)}&=\begin{pmatrix}
0 & 0 & 0 & 0\\
0 & \frac{\sin(x)^2}{2}  & \frac{\cos(x)^2-\sin(x)^2}{2}& \frac{\cos(x)\sin(x)}{2} \\
0 & \frac{\cos(x)^2-\sin(x)^2}{2} & \frac{\sin(x)^2}{2} & \frac{\cos(x)\sin(x)}{2}\\
0 & \frac{\cos(x)\sin(x)}{2} &  \frac{\cos(x)\sin(x)}{2} & \sin(x)^2
\end{pmatrix}
= \begin{pmatrix}
0 & 0 & 0 & 0\\
0 & \frac{1-\cos(2x)}{4} &\frac{cos(2x)}{2} & \frac{\sin(2x)}{4}  \\
0 & \frac{cos(2x)}{2} &\frac{1-\cos(2x)}{4} & \frac{\sin(2x)}{4} \\
0 &  \frac{\sin(2x)}{4}  & \frac{\sin(2x)}{4}  & \frac{1-\cos(2x)}{2}
\end{pmatrix}\\
M^{(11)}&=\begin{pmatrix}
0 & 0 & 0 & 0\\
0 & \frac{\cos(x)^2}{2}  & \frac{\sin(x)^2-\cos(x)^2}{2}& -\frac{\cos(x)\sin(x)}{2} \\
0 & \frac{\sin(x)^2-\cos(x)^2}{2} & \frac{\cos(x)^2}{2} & -\frac{\cos(x)\sin(x)}{2}\\
0 & -\frac{\cos(x)\sin(x)}{2} &  -\frac{\cos(x)\sin(x)}{2} & \cos(x)^2
\end{pmatrix}
= \begin{pmatrix}
0 & 0 & 0 & 0\\
0 & \frac{1+\cos(2x)}{4} & -\frac{cos(2x)}{2} &  -\frac{\sin(2x)}{4}  \\
0 & -\frac{cos(2x)}{2} &\frac{1+\cos(2x)}{4} & - \frac{\sin(2x)}{4} \\
0 &  -\frac{\sin(2x)}{4}  &  -\frac{\sin(2x)}{4}  & \frac{1+\cos(2x)}{2}
\end{pmatrix}.
\end{align}
Note that these POVM matrices are of rank two and can be decomposed into
\beaa \label{decomposition}
M^{(00)}&=\frac{1+3\cos(2x)}{4}\ket{\psi^-}\bra{\psi^-} + \frac{3+\cos(2x)}{4} \ket{\phi^{(00)}}\bra{\phi^{(00)}}
\\
M^{(01)}&=\frac{1-3\cos(2x)}{4}\ket{\psi^-}\bra{\psi^-} + \frac{3-\cos(2x)}{4} \ket{\phi^{(01)}}\bra{\phi^{(01)}}
\\
M^{(10)}&=\frac{1-3\cos(2x)}{4}\ket{\psi^-}\bra{\psi^-} + \frac{3-\cos(2x)}{4} \ket{\phi^{(10)}}\bra{\phi^{(10)}}
\\
M^{(11)}&=\frac{1+3\cos(2x)}{4}\ket{\psi^-}\bra{\psi^-} + \frac{3+\cos(2x)}{4} \ket{\phi^{(11)}}\bra{\phi^{(11)}}, 
\eeaa
with $\ket{\psi^-}= (\ket{01}-\ket{10})/\sqrt{2}$ and where $\ket{\phi^{(ij)}}$ are some tripartite entangled states, given by
\beaa
\ket{\phi^{(00)}} &= \frac{1}{\sqrt{4\cot^2(x)+2}} \left( 2 \cot(x) \ket{00} + \ket{01} + \ket{10} \right)
\\
\ket{\phi^{(01)}} &= \frac{1}{\sqrt{4\tan^2(x)+2}} \left( - 2 \tan(x) \ket{00} + \ket{01} + \ket{10} \right)
\\
\ket{\phi^{(10)}} &= \frac{1}{\sqrt{4\tan^2(x)+2}} \left( 2 \tan(x) \ket{11}+ \ket{01} + \ket{10} \right)
\\
\ket{\phi^{(11)}} &=\frac{1}{\sqrt{4\cot^2(x)+2}} \left( - 2 \cot(x) \ket{11} + \ket{01} + \ket{10} \right).
\eeaa

\subsection{Slightly-coherent unitaries}

Next, we consider the non-perfect cases, i.e., those where
\begin{align}
\sqrt{2} \min\{|U_{10}|^2, |U_{11}|^2\}< |U_{10} U_{11}| \qquad \Leftrightarrow \qquad |U_{10}|^2 < \frac{1}{3}  \quad \Leftrightarrow \quad |U_{10}|^2 > \frac{2}{3}.
\label{case2}
\end{align}
The optimal POVM according to Eq.~\eqref{optmeas} is reached by maximising $|m_1+m_2|$, which is restricted due to positivity of the POVM by \eqref{m1cond} and \eqref{m2cond} and thus maximised by setting $\purple{m_1=m_2=\tilde{m}_1=\tilde{m}_2=\frac{1}{\sqrt{2}} \min \{|U_{10}|^2, |U_{11}|^2\}\frac{U_{10}^\ast U_{11}}{|U_{10}^\ast U_{11}|}}$. Again, we still need to find some $z = \tz'$ and $\tz = z'$, such that all the POVM matrices are positive-semidefinite. Taking $z$ to be real and calculating the determinants of the four matrices, we get the following conditions:
\begin{align}
\left(z - \frac{|U_{10}|^2}{2}\right) \left(z - \frac{2 \min\{|U_{10}|^4, |U_{11}|^4\}-|U_{10}|^4}{2|U_{10}|^2}\right) &\leq 0 \label{firstcond2}
\\
\left(\tz - \frac{|U_{10}|^2}{2}\right) \left(\tz - \frac{2 \min\{|U_{10}|^4, |U_{11}|^4\}-|U_{10}|^4}{2|U_{10}|^2}\right) &\leq 0 \label{secondcond2}
\\
\left(z' + \frac{|U_{11}|^2}{2}\right) \left(z' + \frac{2 \min\{|U_{10}|^4, |U_{11}|^4\}-|U_{11}|^4}{2|U_{11}|^2}\right) &\leq 0 \label{thirdcond2}
\\
\left(\tz' + \frac{|U_{11}|^2}{2}\right) \left(\tz'+  \frac{2 \min\{|U_{10}|^4, |U_{11}|^4\}-|U_{11}|^4}{2|U_{11}|^2}\right) &\leq 0 
\end{align}
We can see that all conditions are satisfied if we take $\orange{z = \tz = z' = \tz' = \frac{|U_{10}|^2}{2}}$ in the case where $|U_{10}|^2 < \frac{1}{3}$ and $\orange{z = \tz = z' = \tz' = -\frac{|U_{11}|^2}{2}}$ in the case where $|U_{11}|^2 < \frac{1}{3}$. The back-action quantified by the difference in final probabilities is proportional to the initial coherence $\rho_{01}$, 
\begin{align}
p_1^{(2)} - p_{01} - p_{11} &= 2 Re[\rho_{01}( U_{10}^\ast U_{11} - \sqrt{2} \min\{U_{10}|^2, |U_{11}|^2\})].\label{difffinalprob}
\end{align}
Note that also in this case, due to $z = z'$, we get the correct initial probabilities and therefore the back-action in terms of difference in average work equals \eqref{difffinalprob}.

Characterizing the unitary again by $U = \sigma_{\mathbbm{z}} \cos(x) + \sigma_{\mathbbm{x}} \sin(x)$, the POVM matrices for $\sin(x)^2<\frac{1}{3}$ (w.l.o.g. we only consider this case, as $\cos(x)^2<\frac{1}{3}$ gives the same back-action) are given by
\begin{align}
M^{(00)}&=\begin{pmatrix}
\cos^2(x) & \frac{\sin^2(x)}{\sqrt{2}} &  \frac{\sin^2(x)}{\sqrt{2}}  & 0\\
\frac{\sin^2(x)}{\sqrt{2}}  & \frac{\cos^2(x)}{2} &  \frac{-\sin^2(x)}{2}  & 0\\
\frac{\sin^2(x)}{\sqrt{2}}  &  \frac{-\sin^2(x)}{2} & \frac{\cos^2(x)}{2} & 0\\
0 & 0 & 0 & 0
\end{pmatrix}\\
M^{(01)}&=\begin{pmatrix}
\sin^2(x) & -\frac{\sin^2(x)}{\sqrt{2}} & -\frac{\sin^2(x)}{\sqrt{2}} & 0\\
- \frac{\sin^2(x)}{\sqrt{2}}  & \frac{\sin^2(x)}{2} & \frac{\sin^2(x)}{2}  & 0\\
- \frac{\sin^2(x)}{\sqrt{2}}  &  \frac{\sin^2(x)}{2}  & \frac{\sin^2(x)}{2} & 0\\
0 & 0 & 0 & 0
\end{pmatrix}\\
M^{(10)}&=\begin{pmatrix}
0 & 0 & 0 & 0\\
0 & \frac{\sin^2(x)}{2}  & \frac{\sin^2(x)}{2} &  \frac{\sin^2(x)}{\sqrt{2}}  \\
0 &  \frac{\sin^2(x)}{2} & \frac{\sin^2(x)}{2} &  \frac{\sin^2(x)}{\sqrt{2}} \\
0 &  \frac{\sin^2(x)}{\sqrt{2}}  &   \frac{\sin^2(x)}{\sqrt{2}}  & \sin^2(x)
\end{pmatrix}\\
M^{(11)}&=\begin{pmatrix}
0 & 0 & 0 & 0\\
0 & \frac{\cos^2(x)}{2}  & \frac{-\sin^2(x)}{2} & - \frac{\sin^2(x)}{\sqrt{2}}  \\
0 &  \frac{-\sin^2(x)}{2}  & \frac{\cos^2(x)}{2} & - \frac{\sin^2(x)}{\sqrt{2}} \\
0 & - \frac{\sin^2(x)}{\sqrt{2}}  &  - \frac{\sin^2(x)}{\sqrt{2}}  & \cos^2(x)
\end{pmatrix}.
\end{align}

\section{Correlated but separable POVMs do no better than factorized POVMs}

In this section, we show that (possibly) correlated but \textit{not} entangled POVMs, which are called separable POVMs, are no better at reducing QBA than factorized, non-correlated POVMs studied in Refs.~\cite{wu2019, perarnau2017nogo}. Therefore, the optimal QBA-reducing POVMs obtained in Supplementary Section \ref{app:newconstruction}, being more optimal than the separable POVMs, are necessarily entangled, thereby proving that entanglement is \textit{the} resource behind the significant QBA reduction reported in this work as compared to previous work.

First, let us specify how to measure entanglement of a POVM. To do so, we notice that each element $M$ of a POVM is a positive-semidefinite Hermitian operator which, in a finite-dimensional Hilbert space, has a finite trace. Therefore $M / \Tr M$ is a density operator, which means that we can define and measure entanglement of a POVM element in the same way as we do for quantum states \cite{virmani2003}. Here, we will measure the entanglement of the POVM as a whole as the arithmetic mean of individual entanglements of the POVM elements: for our case of $4$-element POVM, we thus have
\bea
\mathcal{E}(\mathcal{M}) = \frac{1}{4} \sum_{i, j} \mathcal{E}(M^{(ij)}).
\eea
What is more, since our system of interest merely consists of two qubits, entanglement can be characterized through the so-called positive partial transpose criterion \cite{horodecki2009}. In a given basis, the partial transpose of an operator $M$ with respect to the second particle, $M^{\Gamma_2}$, is defined as $\bra{i_1} \bra{i_2} M^{\Gamma_2} \ket{j_1} \ket{j_2} = \bra{i_1} \bra{j_2} M \ket{j_1} \ket{i_2}$, and the positive partial transpose criterion states that, for $2 \times 2$ and $2 \times 3$ dimensional systems, $M$ is separable if and only if $M^{\Gamma_2}$ is positive-semidefinite \cite{horodecki2009}. (One can of course also consider the partial transpose with respect to the first particle, but it will yield identical results \cite{horodecki2009}.) Moreover, (only) for $2 \times 2$-dimensional systems, as long as $M \geq 0$, $M^{\Gamma_2}$ can have only one negative eigenvalue \cite{sanpera1998, verstraete2001}, and thus $M$'s entanglement can be measured by the absolute value of that negative eigenvalue; accordingly, if all the eigenvalues of the partial transpose of $M$ are nonnegative, $M$ is separable ($\Leftrightarrow$ non-entangled). Lastly, since only one eigenvalue can be negative, the sign of $\det M^{\Gamma_2}$ can also be used to detect entanglement (or certify separability), and we are going to make use of that fact.

\subsubsection{Factorized POVM elements}

Let us determine when $M^{(ij)}$'s in Eq.~\eqref{construction} are factorized (i.e., of the $M_1^{(ij)} \otimes M_2^{(ij)}$ form, where $M_k^{(ij)} \geq 0$). Provided $U_{11} \neq 0$, because otherwise $M^{(00)}$ and $M^{(11)}$ would be $0$, $M^{(00)}$ and $M^{(11)}$ cannot be factorized if $0 < \widetilde{a} < 1$ because, for such values of $\widetilde{a}$, the upper-left and lower-right $2 \times 2$ submatrices of $M^{(00)}$ and $M^{(11)}$ have different ranks. The same argument holds for $M^{(01)}$ and $M^{(10)}$, meaning that there are only four possibilities for $\mathcal{M}$ to be factorized: $(a, \ta) = (0, 0), (0, 1), (1, 0), (1, 1)$, and it is straightforward to see that $m_1$ and $m_2$ cannot be simultaneously nonzero, and similarly, so cannot $\tm_1$ and $\tm_2$. Moreover, the $M^{(ij)} \geq 0$ condition leads us to
\beaa \label{rex}
|m_1| + |m_2| \leq \min \{ |U_{11}|^2, |U_{10}|^2 \},
\\
|\tm_1| + |\tm_2| \leq \min \{ |U_{11}|^2, |U_{10}|^2 \}.
\eeaa
This case has already been studied in Ref.~\cite{wu2019} (see also Ref.~\cite{perarnau2017nogo}).

Lastly, let us note that the probability distribution produced by a factorized POVM $M^{(ij)}$ on $\rho \otimes \rho$, $p_{ij} = \Tr\big(\rho M_1^{(ij)}\big) \Tr\big( \rho M_2^{(ij)} \big)$, can be correlated; see, e.g., the one in Ref.~\cite{perarnau2017nogo}. Instead, the correlations to which we refer throughout this work concern the internal structure of the POVM elements, seen as operators. Namely, if an element of the POVM, $M$, is factorizable, i.e., can be represented as a Kronecker product of two positive-semidefinite Hermitian operators, $M = M_1 \otimes M_2$, then we may say it is uncorrelated despite the fact it may produce correlated statistics. If it cannot be decomposed into a product, but can be represented as $M = \sum_\kappa M_1^{(\kappa)} \otimes M_2^{(\kappa)}$, where all $M_i^{(\kappa)}$ are positive-semidefinite Hermitian operators, then we say $M$ is classically correlated or, synonymously, separable. When $M$ is not separable, we call it entangled.

\subsubsection{Correlated but separable POVM elements}

Correlated but non-entangled POVM elements in Eq.~\eqref{construction} are thus to be sought in the
\bea \label{cujo}
0 < a < 1 \quad \mathrm{and} \quad 0 < \ta < 1
\eea
range, with the condition of both $M^{(ij)} \geq 0$ and $\left( M^{ij} \right)^{\Gamma_2} \geq 0$. Starting with the latter, a necessary condition for $\left( M^{ij} \right)^{\Gamma_2} \geq 0$ to hold is $\det \left( M^{ij} \right)^{\Gamma_2} \geq 0$. E.g., for $M^{(00)}$, $\det \left( M^{00} \right)^{\Gamma_2} = - \ta (1 - \ta) |U_{11}|^4 |x'|^2$, therefore, $\det \left( M^{00} \right)^{\Gamma_2} \geq 0$ is possible only if $z' = 0$. Similarly, the nonneagativity conditions for the partial transposes of the other POVM elements further lead to $z = \tz = \tz' = 0$. With all the $z$'s being $0$, it is straightforward to check all principal minors of $M^{(ij)}$ and show (according to Sylvester's criterion) that the simultaneous satisfaction of $M^{(ij)} \geq 0$ is equivalent to the simultaneous satisfaction of
\beaa \label{chewbacca}
\ta |m_2|^2 + (1 - \ta) |m_1|^2 &\leq \ta (1 - \ta) |U_{11}|^4,
\\
a |m_2|^2 + (1 - a) |m_1|^2 &\leq a (1 - a) |U_{10}|^4,
\\
a |\tm_2|^2 + (1 - a) |\tm_1|^2 &\leq a (1 - a) |U_{10}|^4,
\\
\ta |\tm_2|^2 + (1 - \ta) |\tm_1|^2 &\leq \ta (1 - \ta) |U_{11}|^4;
\eeaa
along with Eq.~\eqref{cujo} and the fact that simultaneously all $\det \left( M^{(ij)} \right)^{\Gamma_2} = 0$ ($z = \tz = z' = \tz' = 0$), Eq.~\eqref{chewbacca} fully characterizes the set of correlated but separable POVMs that satisfy the condition \textcolor{red}{$C_1$} in the main text (i.e., those given by Eq.~\eqref{construction}).

Now, noticing that
\beaa \nonumber
|m_1| + |m_2| &= \frac{1}{\sqrt{\ta}} \cdot \sqrt{\ta} |m_1| + \frac{1}{\sqrt{1 - \ta}} \cdot \sqrt{1 - \ta} |m_2|
\\
&\leq \sqrt{\left( \frac{1}{\sqrt{\ta}} \right)^2 + \left( \frac{1}{\sqrt{1 - \ta}} \right)^2} \sqrt{\ta |m_2|^2 + (1 - \ta) |m_1|^2}
\\
&\leq \sqrt{\frac{1}{\ta (1 - \ta)}} \sqrt{\ta (1 - \ta) |U_{11}|^4} = |U_{11}|^2,
\eeaa
where the first inequality is simply the Cauchy-Schwartz inequality, and performing this for the rest of the lines in Eq~\eqref{chewbacca}, we see that the separability of $\mathcal{M}$ leads exactly to Eq.~\eqref{rex}. In other words, extending the set of allowed POVMs from factorized to separable gives no advantage in reducing QBA, which means that the most optimal factorized POVM already found in Refs.~\cite{wu2019, perarnau2017nogo} is in fact the most optimal separable QBA-reducing POVM.

\section{Experimental aspects}

\subsection{State preparation}\label{app:statepreparation}

We first describe the state preparation module.
Note that the action of a HWP with rotation angle $\theta$ is implementing a unitary transformation on polarization-encoded states, 
\begin{equation}\label{hwptr}
\begin{aligned}
u(\theta)=\begin{pmatrix}
&\cos 2\theta&\sin 2\theta\\
&\sin 2\theta&-\cos 2\theta
\end{pmatrix}.
\end{aligned}
\end{equation}
From Eq.~\eqref{hwptr} we have 
\begin{equation}
\begin{aligned}
&u(0^\circ)=\sigma_{\mathbbm{z}},\,u(90^\circ)=-\sigma_{\mathbbm{z}},\\
&u(22.5^\circ)=\frac{1}{\sqrt{2}}(\sigma_{\mathbbm{x}}+\sigma_{\mathbbm{z}}),\,u(45^\circ)=\sigma_{\mathbbm{x}}\\
&u(90^\circ)u(22.5^\circ)u(0^\circ)=\frac{1}{\sqrt{2}}(\sigma_{\mathbbm{x}}-\sigma_{\mathbbm{z}}),
\end{aligned}
\end{equation}

The state preparation module is shown in detail in Fig.~\ref{fig:exp2}, we use the same technology as in~\cite{wu2019} (Ref. [12] in the main text), and initialize the state as an identically prepared two-copy pure state $\ket{\Phi}\otimes\ket{\Phi}$, where
\begin{equation}\label{pure1}
\ket{\Phi}=\cos\alpha\ket{0}+\sin\alpha\ket{1}.
\end{equation}

In particular, we take advantage of the multiple degrees (polarization and spatial mode) of a single photon. Initially, a single-photon $\ket{H}$ is generated. It passes through a H$_1$ with a rotation angle $\frac{\alpha}{2}$, resulting in a pure polarization-encoded qubit state in Eq.~\eqref{pure1}. Then, the photon passes the beam displacer (BD), the $H$ component is displaced into path $1$, which is 4 mm away from the $V$ component in path $-1$, resulting in a path-polarized entangled state:
\begin{equation}
\cos\alpha\ket{1}\otimes\ket{H}+\sin\alpha\ket{-1}\otimes\ket{V}.
\end{equation}
Following, another fixed half-wave plate with a rotation angle of $45^{\circ}$ flips the $V$-polarized photon to a $H$-polarized photon, resulting in a product state: 
\begin{equation}
\ket{\Phi}\otimes\ket{H}, 
\end{equation}
where $\ket{\Phi}$ denotes the state in Eq.~(\ref{pure1}). Finally, the second copy is encoded into the polarization degree by setting a third half-wave plate to $\frac{\alpha}{2}$ that acts on both paths, thus generating the desired state $\ket{\Phi}\otimes\ket{\Phi}$. Note, that we have taken $0\equiv \{H, \, 1\}$ and $1\equiv \{V, \, -1\}$.

Thus, we can prepare initial states with various coherence $C(\ket{\Phi})=\sin 2\alpha$ (quantified by the sum of the absolute of the off-diagonal elements).

\subsection{Experimental entangled measurement based on photonic quantum walks}\label{app:mreasurement}

\begin{figure}[htp]
	\label{fig:exp2}
	\includegraphics[scale=0.055]{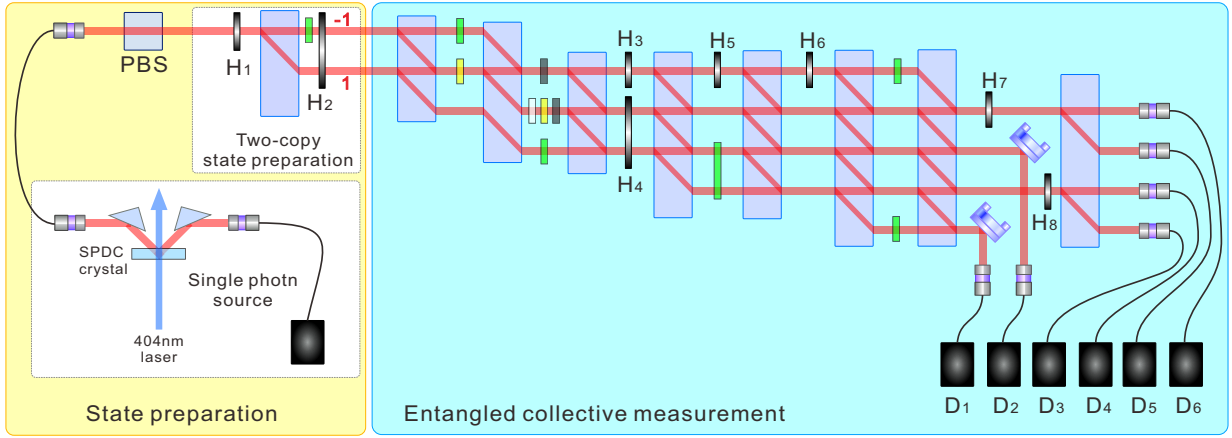}
	\caption{\label{fig:exp2} \textbf{Detail experimental setup.} Details of the whole experimental setup is shown. Yellow area denotes two-copy state preparation module, and blue area denotes entangled collective measurement area. Key to components: SPDC, spontaneous parametric down conversion; PBS, polarizing beam splitter.
	}
\end{figure}

In our previous work \cite{wu2019}, the implemented POVM consisted of four product local measurement operators acting on two degrees of freedom of a single photon. Thus the implementation demanded no interference between photons in each mode. In particular, the measurement acting on the first copy  corresponded to a standard projective measurement and the measurement on the second copy depended on the results of the projective measurement, but could be easily implemented through a standard path measurement and subsequent polarization measurement.

In this work, however, the four entangled POVM operators are highly non-trivial---they are not simply constitute a Bell measurement, which has been experimentally demonstrated before \cite{hou2018}, but they are a convex combination of some non-trivial entangled states, complicating their design and implementation based on current photonic technologies.

Our experimental setup for the entangled measurement is based on a photonic quantum walk \cite{hou2018} (Ref. [55] in the main text; see also Fig.~\ref{fig:exp2}). According to Ref.~\cite{li2019implementation}, any qudit POVM with $n$ rank-$1$ operators can be deterministically implemented via a $2(n-1)$-step one dimensional discrete quantum walk. The optimal POVM for reducing back action contains four non-trivial entangled operators whose ranks sum up to $8$, which would therefore correspond to a $14$-step quantum walk. This would result in a substantial decrease in visibility due to non-ideally identically fabricated beam displacers.

Here, we realize this POVM via a much simpler quantum walk design, taking both the experimental errors and simplicity into consideration. In particular, we experimentally realized a $7$-step quantum walk for the perfect case and an $8$-step quantum walk for the imperfect case, which needs better interference between different spatial modes and polarization of a single photon than what has been reported in \cite{hou2018}. Generally, in a one-dimensional discrete quantum walk, the system is characterized by two degrees of freedom: $p$ and $c$, where $p = ... -1, 0, 1, ...$ denotes the walker position and $c \in \{H, V\}$ represents the coin state. The dynamics of the walker-coin state at each time $t$ can be described by a unitary transformation 
\begin{equation}
U(t) = T_p C(p, t),
\end{equation}
where 
\begin{equation}
\begin{aligned}
&T_p = \sum_{p}\ketbra{p+1, H}{p, H}+\ketbra{p-1, V}{p, V},\\
&C(p, t) = \sum_{p}\ketbra{p}{p}\otimes\mathcal{O}_{p}^t,
\end{aligned}
\end{equation}
and $\mathcal{O}_p^t$ is a site-dependent coin operator.

\subsubsection{Highly coherent case}

Let us first describe the realization of the perfect measurement. The four rank-2 POVMs are deterministically realized as shown in Fig.1 in maintext. The conceptual designing is shown in Fig.1 (a) in maintext. The coin operators are shown in Table.~\ref{tab:perfect} and the corresponding quantum state after each step is shown in Table.~\ref{tab:perfectstate}. In this case, there are four output ports in total, each corresponding to a specific measurement outcome $ij$ generated by the optimal POVM. Let us denote the angle of \orange{H}$_i$ as $\theta_i$, then we have   $\cos^22\theta_3=\frac{3}{2}\sin^2x$, $\theta_{4}=\frac{x}{2}$, $\cos^22\theta_{5}=\frac{1}{3\sin^2x}$, and $\cos^22\theta_{6}=\frac{1-3\cos 2x}{2}$.

\begin{table*}[htp!]\label{tab:perfect}
	\begin{tabular}{c|cccccccc}
		\hline
		\hline
		$(p, 1)$& (-1, 1)& (1, 1)\\
		\hline
		& $I_2$ & $I_2$ \\
		\hline
		\hline
		$(p, 2)$& (-2, 2)& (0, 2)& (2, 2)\\
		\hline
		& $\sigma_{\mathbbm{x}}$ & $\frac{1}{\sqrt{2}}(\sigma_{\mathbbm{z}}+\sigma_{\mathbbm{x}})$& $I_2$  \\
		\hline
		\hline
		$(p, 3)$& (-1, 3)& (1, 3)& (3, 3)\\
		\hline
		& $-\sigma_{\mathbbm{z}}$ & $\frac{1}{\sqrt{2}}(\sigma_{\mathbbm{x}}-\sigma_{\mathbbm{z}})$& $\sigma_{\mathbbm{x}}$ \\
		\hline
		\hline
		$(p, 4)$& (-2, 4)& (0, 4)& (2, 4)\\
		\hline
		& $\sqrt{\frac{3}{2}}\sin x\sigma_{\mathbbm{z}}+\sqrt{1-\frac{3}{2}\sin^2x}\sigma_{\mathbbm{x}}\quad$ & $\cos x\sigma_{\mathbbm{z}}+\sin x\sigma_{\mathbbm{x}}\quad$ & $\cos x\sigma_{\mathbbm{z}}+\sin x\sigma_{\mathbbm{x}}\quad$ \\
		\hline
		\hline
		$(p, 5)$& (-3, 5)& (-1, 5)& (1, 5)& (3, 5)\\
		\hline
		& $\frac{1}{\sqrt{3}\sin x}\sigma_{\mathbbm{z}}+\sqrt{1-\frac{1}{3\sin^2x}}\sigma_{\mathbbm{x}}$ & $I_2$ & $\sigma_{\mathbbm{x}}$ & $\sigma_{\mathbbm{x}}$ \\
		\hline
		\hline
		$(p, 6)$& (-4, 6)& (-2, 6)& (0, 6)& (2, 6)\\
		\hline
		& $\sqrt{\frac{1-3\cos 2x}{2}}\sigma_{\mathbbm{z}}+\sqrt{\frac{1+3\cos 2x}{2}}\sigma_{\mathbbm{x}}$ & $I_2$ & $I_2$ & $I_2$ \\
		\hline
		\hline
		$(p, 7)$& (-5, 7)& (-3, 7)& (-1, 7)& (1, 7)&$\quad$(3, 7)\\
		\hline
		& $\sigma_{\mathbbm{x}}$ & $I_2$ & $I_2$ & $I_2$ & $\quad\sigma_{\mathbbm{x}}$ \\
		\hline
		\hline
	\end{tabular}
	\caption{\label{tab:perfect} Coin operators for the perfect case}
\end{table*}

\begin{table*}[htp!]\label{tab:perfectstate}
	\begin{tabular}{c|cccccccc}
		\hline
		\hline
		Step $t$&Quantum state $\ket{\Phi_t}$\\
		\hline
		\hline
		1& $a\ket{2, H}+b\ket{0, V}+c\ket{0, H}+d\ket{-2, V}$\\
		\hline
		2& $a\ket{3, H}+\frac{b+c}{\sqrt{2}}\ket{1, H}-\frac{b-c}{\sqrt{2}}\ket{-1, V}+d\ket{-1, H}$\\
		\hline
		3& $a\ket{2, V}-\frac{b+c}{2}\ket{2, H}+\frac{b+c}{2}\ket{0, V}-d\ket{0, H}-\frac{b-c}{\sqrt{2}}\ket{-2, V}$\\
		\hline
		4& $\left(a\sin x-\frac{b+c}{2}\cos x\right)\ket{3, H}-\left(a\cos x+\frac{b+c}{2}\sin x\right)\ket{1, V}-\left(\frac{b+c}{2}\cos x+d\sin x\right)\ket{-1, V}+\left(\frac{b+c}{2}\sin x-d\cos x\right)\ket{1, H}$\\
		&$+\sqrt{\frac{3}{2}}\sin x\frac{b-c}{\sqrt{2}}\ket{-3, V}-\sqrt{1-\frac{3}{2}\sin^2x}\frac{b-c}{\sqrt{2}}\ket{-1, H}$\\
		\hline
		5& $\left(a\sin x-\frac{b+c}{2}\cos x\right)\ket{2, V}-\left(a\cos x+\frac{b+c}{2}\sin x\right)\ket{2, H}-\left(\frac{b+c}{2}\cos x+d\sin x\right)\ket{-2, V}+\left(\frac{b+c}{2}\sin x-d\cos x\right)\ket{0, V}$\\
		&$-\frac{b-c}{2}\ket{-4, V}+\sqrt{3\sin^2x-1}\frac{b-c}{2}\ket{-2, H}-\sqrt{1-\frac{3}{2}\sin^2x}\frac{b-c}{\sqrt{2}}\ket{0, H}$\\
		\hline
		6& $\left(a\sin x-\frac{b+c}{2}\cos x\right)\ket{1, V}-\left(a\cos x+\frac{b+c}{2}\sin x\right)\ket{3, H}-\left(\frac{b+c}{2}\cos x+d\sin x\right)\ket{-3, V}+\left(\frac{b+c}{2}\sin x-d\cos x\right)\ket{-1, V}$\\
		&$+\sqrt{\frac{1-3\cos 2x}{2}}\frac{b-c}{2}\ket{-5, V}-\sqrt{\frac{1+3\cos 2x}{2}}\frac{b-c}{2}\ket{-3, H}+\sqrt{3\sin^2x-1}\frac{b-c}{2}\ket{-1, H}-\sqrt{1-\frac{3}{2}\sin^2x}\frac{b-c}{\sqrt{2}}\ket{1, H}$\\
		\hline
		7& $\left(a\sin x-\frac{b+c}{2}\cos x\right)\ket{0, V}-\left(a\cos x+\frac{b+c}{2}\sin x\right)\ket{2, V}-\left(\frac{b+c}{2}\cos x+d\sin x\right)\ket{-4, V}+\left(\frac{b+c}{2}\sin x-d\cos x\right)\ket{-2, V}$\\
		&$+\sqrt{\frac{1-3\cos 2x}{2}}\frac{b-c}{2}\ket{-4, H}-\sqrt{\frac{1+3\cos 2x}{2}}\frac{b-c}{2}\ket{-2, H}-\sqrt{3\sin^2x-1}\frac{b-c}{2}\ket{0, H}-\sqrt{1-\frac{3}{2}\sin^2x}\frac{b-c}{\sqrt{2}}\ket{2, H}$\\
		
		\hline
		\hline
	\end{tabular}
	\caption{\label{tab:perfectstate} Quantum state after each step for the perfect case, assuming an initially pure state $\ket{\Phi_0}=a\ket{1, H}+b\ket{1, V}+c\ket{-1, H}+d\ket{-1, V}$. }
\end{table*}

Assuming the initial hybrid quantum state is a pure state
\begin{equation}\label{initialtwoarbitrary}
\ket{\Phi_0}=a\ket{1, H}+b\ket{1, V}+c\ket{-1, H}+d\ket{-1, V}.
\end{equation}
Then, after the first step, where there is an identity coin operator on the polarization state and a site-dependent translation operator $T$, the quantum state will be, 
\begin{equation}
\ket{\Phi_1}=d\ket{-2, V}+b\ket{0, V}+c\ket{0, H}+a\ket{2, H}
\end{equation}
Following the representation of the coin operators in Tab.~\ref{tab:perfect}, we can derive the quantum state after 7 steps is 

\begin{equation}
\begin{aligned}
\ket{\Phi_7} =& \left(a\sin x-\frac{b+c}{2}\cos x\right)\ket{0, V} - \left(a\cos x+\frac{b+c}{2}\sin x\right)\ket{2, V}
\\
&- \left(\frac{b+c}{2}\cos x+d\sin x\right)\ket{-4, V}+\left(\frac{b+c}{2}\sin x-d\cos x\right)\ket{-2, V}
\\
&+\sqrt{\frac{1-3\cos 2x}{2}}\frac{b-c}{2}\ket{-4, H}-\sqrt{\frac{1+3\cos 2x}{2}}\frac{b-c}{2}\ket{-2, H}
\\
&+\sqrt{3\sin^2x-1}\frac{b-c}{2}\ket{0, H}-\sqrt{1-\frac{3}{2}\sin^2x}\frac{b-c}{\sqrt{2}}\ket{2, H}
\end{aligned}
\end{equation}

Then, at the end of the quantum walk, we can experimentally determine the probability for a photon to be in each position, i.e., $x = -4, -2, 0$ or $2$. We have

\begin{equation}
\begin{aligned}
\mathrm{Prob}(x=-4) &= \left|\frac{b+c}{2}\cos x+d\sin x\right|^2+\frac{1-3\cos 2x}{8}|b-c|^2,
\\
\mathrm{Prob}(x=-2) &= \left|\frac{b+c}{2}\sin x-d\cos x\right|^2+\frac{1+3\cos 2x}{8}|b-c|^2,
\\
\mathrm{Prob}(x=0) &= \left|a\sin x-\frac{b+c}{2}\cos x\right|^2+\frac{1-3\cos 2x}{8}|b-c|^2,
\\
\mathrm{Prob}(x=2) &= \left|a\cos x+\frac{b+c}{2}\sin x\right|^2+\frac{1+3\cos 2x}{8}|b-c|^2.
\end{aligned}
\end{equation}

In our experiments, we use the same setup for both cases. In the perfect case, we obtain the work distribution produced by the measurement on the state after 7 steps, which is given by the probability of obtaining a single photon coincident event from the output port of the quantum walk device, i.e.,
\begin{equation}
\begin{aligned} \label{probabilityeq2}
\mathrm{Prob}(x=-4) &= \Tr\left(M^{(10)}\ketbra{\Phi_0}{\Phi_0}\right) = p_{10}=\frac{C_{D_5}+C_{D_6}}{\sum_iC_{D_i}},
\\
\mathrm{Prob}(x=-2) &= \Tr\left(M^{(11)}\ketbra{\Phi_0}{\Phi_0}\right) = p_{11}=\frac{C_{D_2}}{\sum_iC_{D_i}},
\\
\mathrm{Prob}(x=0) &= \Tr\left(M^{(01)}\ketbra{\Phi_0}{\Phi_0}\right) = p_{01}=\frac{C_{D_3}+C_{D_4}}{\sum_iC_{D_i}},
\\
\mathrm{Prob}(x=2) &= \Tr\left(M^{(00)}\ketbra{\Phi_0}{\Phi_0}\right) = p_{00}=\frac{C_{D_1}}{\sum_iC_{D_i}},
\end{aligned}
\end{equation}
where $C_{D_i}$ denotes the photon counting from $i$'th detector. Here, we just use the state after $7$ steps, as the last two half-wave plates, $H_7$ and $H_8$, are set to $0^\circ$. To evaluate the probabilities produced by $M^{(10)}$ and $M^{(01)}$ we use the combined outcome of two detectors each.

Note, that any mixed states can be written as a linear combination of pure states, so Eq.~(\ref*{probabilityeq2}) is also valid for mixed inputs.

\begin{table*}[htp!]\label{tab:imperfect}
	\begin{tabular}{c|cccccccc}
		\hline
		\hline
		$(p, 1)$& (-1, 1)& (1, 1)\\
		\hline
		& $I_2$ & $I_2$ \\
		\hline
		\hline
		$(p, 2)$& (-2, 2)& (0, 2)& (2, 2)\\
		\hline
		& $\sigma_{\mathbbm{x}}$ & $\frac{1}{\sqrt{2}}(\sigma_{\mathbbm{z}}+\sigma_{\mathbbm{x}})$ & $I_2$ \\
		\hline
		\hline
		$(p, 3)$& (-1, 3)& (1, 3)& (3, 3)\\
		\hline
		& $-\sigma_{\mathbbm{z}}$  & $\frac{1}{\sqrt{2}}(\sigma_{\mathbbm{x}}-\sigma_{\mathbbm{z}})$& $\sigma_{\mathbbm{x}}$ \\
		\hline
		\hline
		$(p, 4)$& (-2, 4)& (0, 4)& (2, 4)\\
		\hline
		& $\frac{1}{\sqrt{2}}(\sigma_{\mathbbm{z}}+\sigma_{\mathbbm{x}})$ & $\sqrt{\frac{2}{3}}\sigma_{\mathbbm{z}}+\sqrt{\frac{1}{3}}\sigma_{\mathbbm{x}}$ & $\sqrt{\frac{2}{3}}\sigma_{\mathbbm{z}}+\sqrt{\frac{1}{3}}\sigma_{\mathbbm{x}}$ \\
		\hline
		\hline
		$(p, 5)$& (-3, 5)& (-1, 5)& (1, 5)& (3, 5)\\
		\hline
		& $\sigma_{\mathbbm{z}}$ & $I_2$ & $\sigma_{\mathbbm{x}}$ & $\sigma_{\mathbbm{x}}$ \\
		\hline
		\hline
		$(p, 6)$& (-4, 6)& (-2, 6)& (0, 6)& (2, 6)\\
		\hline
		& $\sigma_{\mathbbm{x}}$ & $I_2$ & $I_2$ & $I_2$ \\
		\hline
		\hline
		$(p, 7)$& (-5, 7)& (-3, 7)& (-1, 7)& (1, 7)&$\quad$(3, 7)\\
		\hline
		& $\sigma_{\mathbbm{x}}$ & $I_2$ & $I_2$ & $I_2$ & $\quad\sigma_{\mathbbm{x}}$ \\
		\hline
		\hline
		$(p, 8)$& (-4, 8)& (0, 8)\\
		\hline
		& $\; \sigma_{\mathbbm{x}} \sqrt{3} \sin x + \sigma_{\mathbbm{z}} \sqrt{1-3\sin^2x} \;$ & $\; \sigma_{\mathbbm{x}} \sqrt{3} \sin x + \sigma_{\mathbbm{z}} \sqrt{1-3\sin^2x} \;$ \\
		\hline
		\hline
	\end{tabular}
	\caption{\label{tab:imperfect} Coin operators for imperfect case}
\end{table*}

\begin{table*}[htp!]\label{tab:imperfectstate}
	\begin{tabular}{c|cccccccc}
		\hline
		\hline
		Step $t$&Quantum state $\ket{\Phi_t}$ \\
		\hline
		\hline
		1& $a\ket{2, H}+b\ket{0, V}+c\ket{0, H}+d\ket{-2, V}$ \\
		\hline
		2& $a\ket{3, H}+\frac{b+c}{\sqrt{2}}\ket{1, H}-\frac{b-c}{\sqrt{2}}\ket{-1, V}+d\ket{-1, H}$ \\
		\hline
		3& $a\ket{2, V}-\frac{b+c}{2}\ket{2, H}+\frac{b+c}{2}\ket{0, V}-d\ket{0, H}-\frac{b-c}{\sqrt{2}}\ket{-2, V}$ \\
		\hline
		4& $\left[\sqrt{\frac{1}{3}}a-\sqrt{\frac{1}{6}}(b+c)\right]\ket{3, H}-\left[\sqrt{\frac{2}{3}}a+\sqrt{\frac{1}{12}}(b+c)\right]\ket{1, V}-\left[\sqrt{\frac{1}{6}}(b+c)+\sqrt{\frac{1}{3}}d\right]\ket{-1, V}+\left[\sqrt{\frac{1}{12}}(b+c)-\sqrt{\frac{2}{3}}d\right]\ket{1, H}$ \\
		&$+\frac{b-c}{2}\ket{-3, V}-\frac{b-c}{2}\ket{-1, H}$ \\
		\hline
		5& $\left[\sqrt{\frac{1}{3}}a-\sqrt{\frac{1}{6}}(b+c)\right]\ket{2, V}-\left[\sqrt{\frac{2}{3}}a+\sqrt{\frac{1}{12}}(b+c)\right]\ket{2, H}-\left[\sqrt{\frac{1}{6}}(b+c)+\sqrt{\frac{1}{3}}d\right]\ket{-2, V}+\left[\sqrt{\frac{1}{12}}(b+c)-\sqrt{\frac{2}{3}}d\right]\ket{0, V}$ \\
		&$-\frac{b-c}{2}\ket{-4, V}-\frac{b-c}{2}\ket{0, H}$ \\
		\hline
		6&  $\left[\sqrt{\frac{1}{3}}a-\sqrt{\frac{1}{6}}(b+c)\right]\ket{1, V}-\left[\sqrt{\frac{2}{3}}a+\sqrt{\frac{1}{12}}(b+c)\right]\ket{3, H}-\left[\sqrt{\frac{1}{6}}(b+c)+\sqrt{\frac{1}{3}}d\right]\ket{-3, V}+\left[\sqrt{\frac{1}{12}}(b+c)-\sqrt{\frac{2}{3}}d\right]\ket{-1, V}$ \\
		&$-\frac{b-c}{2}\ket{-3, H}-\frac{b-c}{2}\ket{1, H}$ \\
		\hline
		7&  $\left[\sqrt{\frac{1}{3}}a-\sqrt{\frac{1}{6}}(b+c)\right]\ket{0, V}-\left[\sqrt{\frac{2}{3}}a+\sqrt{\frac{1}{12}}(b+c)\right]\ket{2, V}-\left[\sqrt{\frac{1}{6}}(b+c)+\sqrt{\frac{1}{3}}d\right]\ket{-4, V}+\left[\sqrt{\frac{1}{12}}(b+c)-\sqrt{\frac{2}{3}}d\right]\ket{-2, V}$ \\
		&$-\frac{b-c}{2}\ket{-2, H}-\frac{b-c}{2}\ket{2, H}$ \\
		\hline
		8& $\left[a-\sqrt{\frac{1}{2}}(b+c)\right]\sin x\ket{1, H}-\left[\sqrt{\frac{1}{3}}a-\sqrt{\frac{1}{6}}(b+c)\right]\sqrt{1-3\sin^2x}\ket{-1, V}$ \\ &$-\left[\sqrt{\frac{2}{3}}a+\sqrt{\frac{1}{12}}(b+c)\right]\ket{2, V}-\left[\sqrt{\frac{1}{2}}(b+c)+d\right]\sin x\ket{-3, H}$ \\
		&$+\left[\sqrt{\frac{1}{6}}(b+c)+\sqrt{\frac{1}{3}}d\right]\sqrt{1-3\sin^2x}\ket{-5, V}+\left[\sqrt{\frac{1}{12}}(b+c)-\sqrt{\frac{2}{3}}d\right]\ket{-2, V}$ \\
		&$-\frac{b-c}{2}\ket{-2, H}-\frac{b-c}{2}\ket{2, H}$ \\
		
		\hline
		\hline
	\end{tabular}
	\caption{\label{tab:imperfectstate} Quantum state after each step for imperfect case, assuming an initially pure state $\ket{\Phi_0}=a\ket{1, H}+b\ket{1, V}+c\ket{-1, H}+d\ket{-1, V}$. }
\end{table*}

\subsubsection{Slightly coherent case}

Moving to the case of slightly coherent unitaries, we take as initial state again the pure state from Eq.~\eqref{initialtwoarbitrary}.

In this case, we need one more step and the coin operators are different. The angles for each wave plate read: $\theta_3=22.5^\circ$, $\theta_5=0^\circ$, $\theta_6=45^\circ$,  $\cos^22\theta_{7}=\cos^22\theta_{8}=1-3\sin^2x$,  and $\cos^22\theta_{4}=\frac{2}{3}$. According to Table.~\ref{tab:imperfect}, \ref{tab:imperfectstate}, the quantum state after 8 steps is

\begin{equation}
\begin{aligned}
\ket{\Phi_8} =& \left[a-\sqrt{\frac{1}{2}}(b+c)\right]\sin x\ket{1, H}-\left[\sqrt{\frac{1}{3}}a-\sqrt{\frac{1}{6}}(b+c)\right]\sqrt{1-3\sin^2x}\ket{-1, V} \\ &-\left[\sqrt{\frac{2}{3}}a+\sqrt{\frac{1}{12}}(b+c)\right]\ket{2, V}-\left[\sqrt{\frac{1}{2}}(b+c)+d\right]\sin x\ket{-3, H} \\
&+\left[\sqrt{\frac{1}{6}}(b+c)+\sqrt{\frac{1}{3}}d\right]\sqrt{1-3\sin^2x}\ket{-5, V}+\left[\sqrt{\frac{1}{12}}(b+c)-\sqrt{\frac{2}{3}}d\right]\ket{-2, V} \\
&-\frac{b-c}{2}\ket{-2, H}-\frac{b-c}{2}\ket{2, H}
\end{aligned}
\end{equation}

Then, at the end of the quantum walk, we can experimentally determine the probability for a photon to be in each position, i.e., $x=-5, -3, -2, -1, 1$ or $2$. We have:
\begin{equation}
\begin{aligned}
\mathrm{Prob}(x=-5) &= \left|\sqrt{\frac{1}{6}}(b+c)+\sqrt{\frac{1}{3}}d\right|^2(1-3\sin^2x),
\\
\mathrm{Prob}(x=-3) &= \left|\sqrt{\frac{1}{2}}(b+c)+d\right|^2\sin^2 x,
\\
\mathrm{Prob}(x=-2) &= \left|\sqrt{\frac{1}{12}}(b+c)-\sqrt{\frac{2}{3}}d\right|^2+\frac{|b-c|^2}{4},
\\
\mathrm{Prob}(x=-1) &=\left|\sqrt{\frac{1}{3}}a-\sqrt{\frac{1}{6}}(b+c)\right|^2(1-3\sin^2x)
\\
\mathrm{Prob}(x=1) &=  \left|a-\sqrt{\frac{1}{2}}(b+c)\right|^2\sin^2 x,
\\
\mathrm{Prob}(x=2) &= \left|\sqrt{\frac{2}{3}}a+\sqrt{\frac{1}{12}}(b+c)\right|^2+\frac{|b-c|^2}{4}.
\end{aligned}
\end{equation}

We can directly verify that the probability distribution produced by the new entangled measurement scheme is given by the probability of obtaining a single photon coincident event from the output port of the quantum walk device, i.e., 

\begin{equation}
\begin{aligned}
\mathrm{Prob}(x=-3) &= \Tr\left(M^{(10)}\ketbra{\Phi_0}{\Phi_0}\right) = p_{10}=\frac{C_{D_5}}{\sum_iC_{D_i}},
\\
\mathrm{Prob}(x=-5) + P(x=-2) &= \Tr\left(M^{(11)}\ketbra{\Phi_0}{\Phi_0}\right) = p_{11}=\frac{C_{D_2}+C_{D_6}}{\sum_iC_{D_i}},
\\
\mathrm{Prob}(x=1) &= \Tr\left(M^{(01)}\ketbra{\Phi_0}{\Phi_0}\right) = p_{01}=\frac{C_{D_3}}{\sum_iC_{D_i}},
\\
\mathrm{Prob}(x=-1) + \mathrm{Prob}(x=2) &= \Tr\left(M^{(00)}\ketbra{\Phi_0}{\Phi_0}\right) = p_{00}=\frac{C_{D_1}+C_{D_4}}{\sum_iC_{D_i}}.
\end{aligned}
\end{equation}
In conclusion, this setup can be used to generate statistics produced by optimal entangled measurement in both, the perfect and the imperfect case.

\subsection{Details on experimental results}\label{app:results}

The fidelity between two POVMs can be defined similarly to the fidelity between quantum states. Indeed, 
\begin{equation}
\mathcal{F}(M, M') = \left(\sum_i w_i \sqrt{\mathcal{F}_i}\right)^2, 
\end{equation}
with
\bea
w_i = \frac{\sqrt{\Tr(M_i) \Tr(M'_i)}}{d}
\eea
where $d$ is the dimension of the Hilbert space and $\mathcal{F}_i$ is the fidelity between the normalized POVM elements $\mu_i := \frac{M_i}{\Tr(M_i)}$ and $\mu'_i := \frac{M'_i}{\Tr(M'_i)}$: $\mathcal{F}_i = \big( \Tr \sqrt{\sqrt{\mu_i} \, \mu'_i \, \sqrt{\mu_i}} \big)^2$ \cite{wiseman2010}. 

In the main text, we have provided the average fidelity of the POVM when we set $U_{01} = 1 / 3$, which is about $0.985$, and the exact fidelity for four POVM elements read $\mathcal{F}_{00} = 0.996$, $\mathcal{F}_{01} = 0.987$, $\mathcal{F}_{10} = 0.980$ and $\mathcal{F}_{11} = 0.982$. We note that the fidelity is slightly lower than in \cite{hou2018}, where the POVM consists of four rank-$1$ product measurement operators and one Bell measurement. The reasons are as follows: (1) the photonic quantum walk needs more steps this time, where more robust interference between photons in different modes is required; (2) our photonic quantum walk consists of four non-trivial entangled measurement operators which are beyond rank-$1$; (3) we need to balance the coupling efficiency for differently polarized photons in one output port, which is relatively hard to realize when four ports need to be balanced simultaneously.

\begin{figure}[htp]
	\label{fig:result4}
	\includegraphics[scale=0.2]{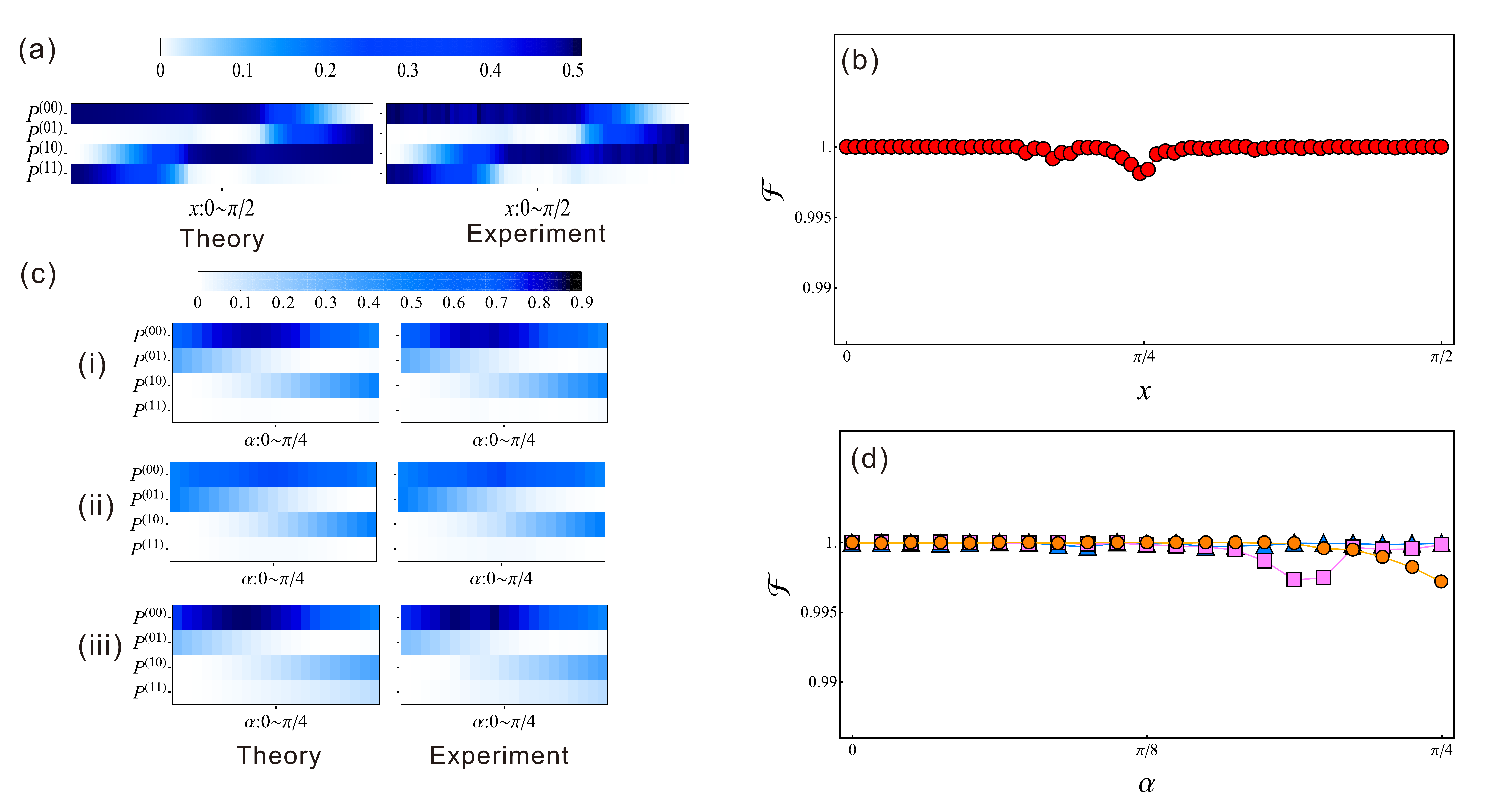}
	\caption{\label{fig:result4} \textbf{Experimental detailed results for different processes and inputs.} \textbf{(a)} Experimental transition probabilities for fixed input $\ket{+}$: theory vs experiment. \textbf{(b)} Experimental fidelities for different processes with fixed input $\ket{+}$, plotted against $x$. \textbf{(c)} Experimental transition probabilities for fixed processes (i) $U_{10} = 1 / \sqrt{3}$, (ii) $U_{10} = 1 / \sqrt{2}$ and (iii) $U_{10} = 1 / 2$: theory vs experiments. \textbf{(d)} Experimental fidelities for different inputs with fixed process, $U_{01} = 1 / 2$ (blue), $1 / \sqrt{3}$ (pink), and $1 / \sqrt{2}$ (orange), plotted against $\alpha$. The fidelity $\mathcal{F} = \mathcal{F}\big(\big\{p_j\big\}, \big\{\widetilde{p}_j\big\}\big)$ is evaluated according to Eqs.~\eqref{karot} and \eqref{tun}, where $p_j$ and $\widetilde{p}_j$ are calculated, respectively, theoretically and experimentally for the entangled measurement scheme.
	}
\end{figure}

The high fidelity of our experimental POVM is also evidenced by Fig.~\ref{fig:result4}, where we show the experimental transition probabilities and fidelities between theoretical predictions and experimental values for our entangled schemes, i.e.,
\bea \label{karot}
\mathcal{F}\big(\big\{p_j\big\}, \big\{\widetilde{p}_j\big\}\big) = \sum_j \sqrt{p_j \widetilde{p}_j},
\eea
where
\bea \label{tun}
p_j := \sum_i p_{ij} = \sum_i \Tr \big( \rho^{\otimes 2} M^{(ij)} \big) \qquad \mathrm{and} \qquad \widetilde{p}_j = \sum_i \widetilde{p}_{ij}.
\eea

From Fig.~\ref{fig:result4}(b), we can see that the lowest fidelity is over $0.998$. A slight drop near $x \sim \pi / 4$ is due to the non-ideal superposition between different modes in the photonic quantum walk, which will add noise into the ideally pure walker-coin state. In particular, the unitary transforms $\ket{+}$ to $\ket{0}$, and ideally we have $\widetilde{p}_0 = 1$ and $\widetilde{p}_1 = 0$. The non-ideal superposition will result in a non-zero $\widetilde{p}_1$ limited by the purity of the overall quantum states, thus the fidelity drops more obviously as the process and initial states are both highly coherent compared to others.

From Fig.~\ref{fig:result4}(d), we can see that the fidelity is beyond $0.997$. Similarly, we observe a decrease in fidelity in the case when the highly-coherent unitary takes a coherent state to an incoherent one, which can also be explained using the aforementioned argument. In contrast, in the slightly coherent case we cannot observe an obvious decrease, which is partly due to the unitary inducing little coherence compared to the other two unitaries.

In our experiments, we mainly consider the statistical errors, since the fluctuation in photon numbers is greater than other systematic error (like uncertainty in rotation angles of wave plates and phase between photons in different modes). The error bars in main text are estimated by Monte Carlo simulation based on the Poisson distribution of photon statistics. In particular, an experimental measured quantity $\widetilde{Q}=f\left(\{\widetilde{n}_i\}\right)$ is determined by a set of data which contains the times of coincident events $\{\widetilde{n}_i\}$. We first generate $N=1000$ data sets $\{\widetilde{n}_i\}_k$ according to the Poisson distribution with average equal to every experimentally obtained coincident events $n_i$, then the sample standard deviation is evaluated by $\widetilde{\sigma}_{Q}:=\sqrt{\sum_{k=1}^{N}\frac{(Q_k-\bar{Q})^2}{N-1}}$, where $Q_k=f\left(\{\widetilde{n}_i\}_k\right)$, and $\bar{Q}$ is the mean value of $\{Q_k\}$.

In our experiments, the probability $\widetilde{p}_{ij}$ for each measurement outcome is calculated as
	\begin{equation}\label{eq:Epro}
	\widetilde{p}_{ij}=\frac{\widetilde{n}_{ij}}{\sum_{ij}\widetilde{n}_{ij}},
	\end{equation}
	where $\widetilde{n}_{ij}$ represents coincident event from the outcome $ij$. With Eq.~(\ref{eq:Epro}), we can evaluate the experimental backaction via $\widetilde{\mathcal{B}} = \sqrt{\sum_j \left(\widetilde{p}_j - p^{(2)}_j \right)^2}$ and $\widetilde{p}_{j}=\sum_i\widetilde{p}_{ij}$.

\end{widetext}

\end{document}